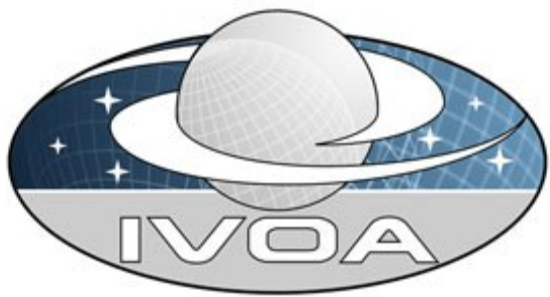

*International*

*Virtual*

*Observatory*

*Alliance*

# Observation Data Model Core Components and its Implementation in the Table Access Protocol

## Version 1.0
### IVOA Recommendation, October 28 2011

**This version:**
> http://www.ivoa.net/Documents/ObsCore/20111028/REC-ObsCore-v1.0-20111028.pdf

**Latest version:**
> http://www.ivoa.net/Documents/ObsCore/20111028/REC-ObsCore-v1.0-20111028.pdf

**Previous version(s):**
> http://www.ivoa.net/Documents/ObsCore/20111008/PR-ObsCore-v1.0-20111008.pdf

**Editors:**
> Doug Tody, Alberto Micol, Daniel Durand, Mireille Louys

**Authors:**
> Mireille Louys, Francois Bonnarel, David Schade, Patrick Dowler, Alberto Micol, Daniel Durand, Doug Tody, Laurent Michel, Jesus Salgado, Igor Chilingarian, Bruno Rino, Juan de Dios Santander, Petr Skoda

## Abstract

This document defines the core components of the Observation data model that are necessary to perform data discovery when querying data centers for observations of interest. It exposes use-cases to be carried out, explains the model and provides guidelines for its implementation as a data access service based on the Table Access Protocol (TAP). It aims at providing a simple model easy to understand and to implement by data providers that wish to publish their data into the Virtual

Observatory. This interface integrates data modeling and data access aspects in a single service and is named ObsTAP. It will be referenced as such in the IVOA registries. There will be a separate document to cover the full Observation data model. In this document, the Observation Data Model Core Components (ObsCoreDM) defines the core components of queryable metadata required for global discovery of observational data. It is meant to allow a single query to be posed to TAP services at multiple sites to perform global data discovery without having to understand the details of the services present at each site. It defines a minimal set of basic metadata and thus allows for a reasonable cost of implementation by data providers. The combination of the ObsCoreDM with TAP is referred to as an ObsTAP service. As with most of the VO Data Models, ObsCoreDM makes use of STC, Utypes, Units and UCDs. The ObsCoreDM can be serialized as a VOTable. ObsCoreDM can make reference to more complete data models such as ObsProvDM (the Observation Provenance Data Model, to come), Characterisation DM, Spectrum DM or Simple Spectral Line Data Model (SSLDM).

# Status of this document

This document has been produced by the IVOA Data Model (DM) working group, in coordination with partners involved in the definition of data access protocols (DAL) and of the ADQL language. It describes the core components of the Observation data model and the metadata to be attached to an astronomical observation, and contains a guide for implementing this model within the Table Access Protocol (TAP) framework. Due to the DM and DAL aspects of this document, this will circulate and be reviewed by both Working Groups. The document content has been worked out as working draft in a previous stage (2009-2010) and is now proposed for IVOA recommendation.

A list of current IVOA Recommendations and other technical documents can be found at *http://www.ivoa.net/Documents/*

# Acknowledgements

This work has been partly funded by Euro-VO AIDA project that we acknowledge here. SSC XMM Catalog service supported the implementation of the SAADA version of ObsTAP at Strasbourg Observatory. The US-VAO project contributed to developing this specification and prototyping the use of ObsTAP in the VAO portal. The CANFAR project also contributed for the reference implementation of ObsTAP at CADC, Victoria.

# Contents

# List of Acronyms

| ADQL | Astronomical Data Query Language |
|---|---|
| DAL | Data Access Layer |
| DM | Data Model |
| ObsCoreDM | Observation Core components Data Model |
| ObsTAP | TAP interface to Observation Data Model |
| TAP | Table Access Protocol |
| SIA | Simple Image Access |
| SSA | Simple Spectral Access |
| STC | Space-Time Coordinates |
| UCD | Unified Content Descriptor |

# 1. Introduction

This work originates from an initiative of the IVOA Take Up Committee that, in the course of 2009, collected a number of use cases for data discovery (see Appendix A). These use cases address the problem of an astronomer posing a world-wide query for scientific data with certain characteristics and eventually retrieving or otherwise accessing selected data products thus discovered. The ability to pose a single scientific query to multiple archives simultaneously is a fundamental use case for the Virtual Observatory. Providing a simple standard protocol such as the one described in this document increases the chances that a majority of the data providers in astronomy will be able to implement the protocol, thus allowing data discovery for almost all archived astronomical observations.

This effort (version 1) is focused on public data. Provision to cover proprietary data is already in preparation (e.g. *obs_release_date* and *data_rights* in the list of optional fields), but is not part of this release. Future versions might cover that in detail.

In the following are described the fundamental building blocks which are used to achieve the goal of *global data discoverability and accessibility*.

## 1.1. First building block: Data Models

Modeling of observational metadata has been an important activity of the IVOA since its creation in 2002. This modeling effort has already resulted in a number of integrated and approved IVOA standards such as the Resource Metadata, Space Time Coordinates (STC), Spectrum and SSA, and the Characterisation data models that are currently used in IVOA services and applications.

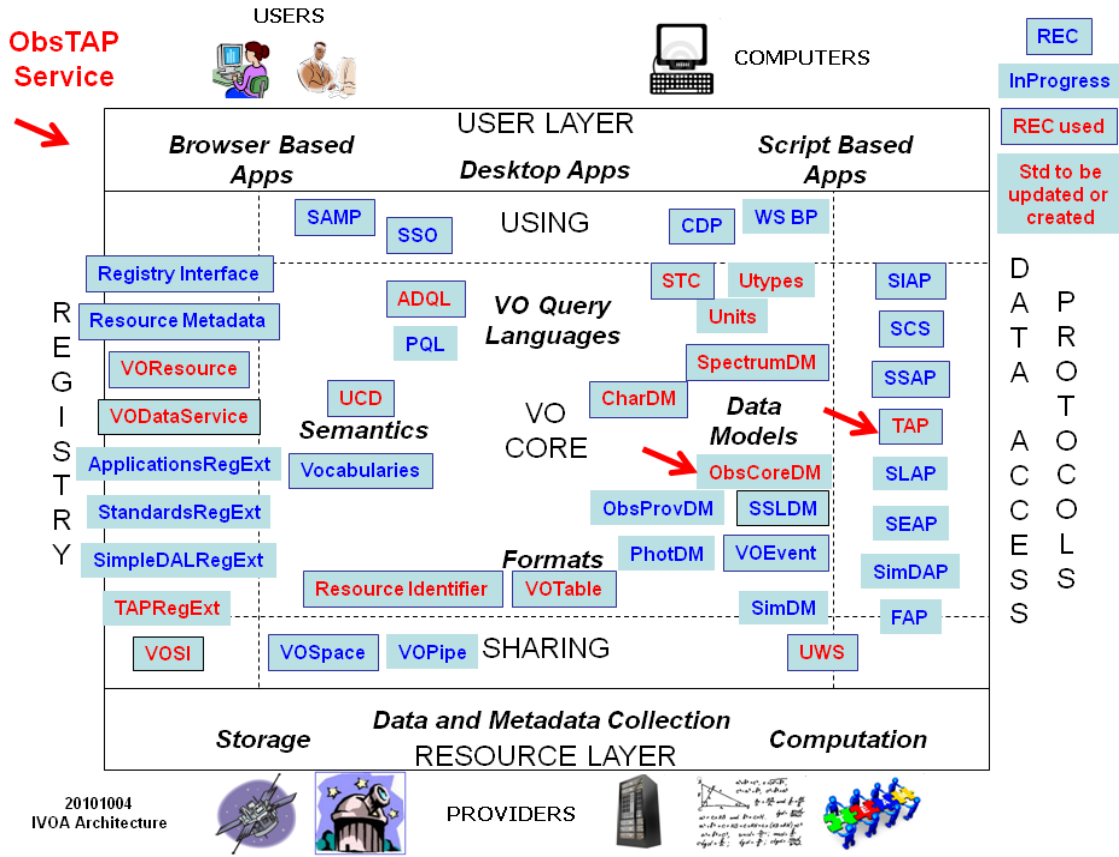

**Figure 1**. How the Observation data model Core Components fits into the overall IVOA architecture. Highlighted blocks in red are data models or specifications that are used by this model.

## 1.2.    Second building block: the Table Access Protocol (TAP)

TAP defines a service protocol for accessing tabular data such as astronomical catalogues, or more generally, database tables. TAP allows a client to (step 1) browse through the various tables and columns (names, units, etc.) in an archive to collect the information necessary to pose a query, then (step 2) actually perform a table query. The Table Access Protocol (TAP) specification was developed and reached recommendation status in March 2010 (Dowler, Tody, & Rixon, 2010).

## 1.3.    The goal of this effort

Building on the work done on data models and TAP, it becomes possible to define a standard service protocol to expose standard metadata describing available datasets. In general, any data model can be mapped to a relational database and exposed directly with the TAP protocol. The goal of ObsTAP is to provide such a capability based upon an essential subset of the general observational data model.

Specifically, this effort aims at defining a database table to describe astronomical datasets (data products) stored in archives that can be queried directly with the TAP protocol. This is ideal for global data discovery as any type of data can be described in a straightforward and uniform fashion. The described datasets can be directly downloaded, or IVOA Data Access Layer (DAL) protocols such as for accessing images (SIA) or spectra (SSA) can be used to perform more advanced data access operations on the referenced datasets.

The final capability required to support uniform global data discovery and access, with a client sending one and the same query to multiple TAP services, is the stipulation that a uniform standard data model is exposed (through TAP) using agreed naming conventions, formats, units, and reference systems. Defining this core data model and associated query mechanism is what this document is for.

Thus the purpose of this document is twofold: (1) to define a simple data model to describe observational data, and (2) to define a standard way to expose it through the TAP protocol to provide a uniform interface to discover observational science data products of any type.

This document is organized as follows:

- Section 2 briefly presents the types of the use cases collected from the astronomical community by the IVOA Uptake committee.
- Section 3 defines the core components of the Observation data model. The elements of the data model are summarized in **Figure 2**. Mandatory ObsTAP fields are summarized in Table 1.
- Section 4 specifies the required data model fields as they are used in the TAP service: table names, column names, column data type, UCD, Utype from the Observation Core components data model, and required units.
- Section 5 describes how to register an ObsTAP service in a Virtual Observatory registry. More detailed information is available in the appendices.
- Examples are cited in section 6
- Section 7 summarizes updates of this document.
- Appendix A describes all the use cases as defined by the IVOA Take Up Committee.
- Appendix B contains a full description of the Observation data model Core Components.
- Appendix C shows the detailed content of the TAP_SCHEMA tables and how to build up and fill them for the implementation of an ObsTAP service.

# 2. Use cases

Our primary focus is on data discovery. To this end a number of use-cases have been defined, aimed at finding observational data products in the VO domain by broadcasting the same query to multiple archives (*global data discoverability and accessibility*). To achieve this we need to give data providers a set of metadata attributes that they can easily map to their database system in order to support queries of the sort listed below.

The goal is to be simple enough to be practical to implement, without attempting to exhaustively describe every particular dataset.

The main features of these use-cases are as follows:

- Support multi-wavelength as well as positional and temporal searches.
- Support any type of science data product (image, cube, spectrum, time series, instrumental data, etc.).
- Directly support the sorts of file content typically found in archives (FITS, VOTable, compressed files, instrumental data, etc.).

Further server-side processing of data is possible but is the subject of other VO protocols. More refined or advanced searches may include extra knowledge obtained by prior queries to determine the range of data products available.

The detailed list of use cases proposed for data discovery is given in Appendix A.

# 3. Observation Core Components Data Model

This section highlights and describes the *core components* of the Observation data model. The term "core components" is meant to refer to those elements of the larger Observation Data Model that are required to support the use cases listed in Appendix A.  In reality this effort is the outcome of a trade-off between what astronomers want and what data providers are ready to offer.  The aim is to achieve buy-in of data providers with a simple and "good enough" model to cover the majority of the use cases.

The project of elaborating a general data model for the metadata necessary to describe any astronomical observation was launched at the first Data Model WG meeting held in Cambridge, UK at the IVOA meeting in May 2003. The Observation data model was sketched out relying on some key concepts: Dataset, Identification, Curation, physical Characterisation and Provenance (either instrumental or software).  A description of the early stages of this development can be found in (Mc Dowell & al., 2005) (Observation IVOA note).  Some of these concepts have already been elaborated in existing data models, namely the Spectrum data model (McDowell, Tody, & al, 2011) for general items such as dataset identification and curation, and the Characterisation data model (Louys & DataModel-WG., 2008) for the description of the physical axes and properties of an observation, such as coverage, resolution, sampling, and accuracy.  The Core Components data model reuses the relevant elements from those models.  Generalization of the observational model to support data from theoretical models (e.g., synthetic spectra) is possible but is not addressed here in order to keep the core model simple.

## 3.1.   UML description of the model

This section provides a graphical overview of the Observation Core Components data model using the unified modeling language (UML).  The UML class diagram shown in Figure 2 depicts the overall Observation Data Model, detailing those aspects that are relevant to the Core Components, while omitting those not relevant.  The Characterisation classes describing how the data span along the main physical measurement axes are simplified here showing only the attributes necessary for data discovery.   This is also the case for the DataID and Curation classes extracted from the Spectrum/SSA data model where only a subset of the attributes are actually necessary for data discovery.  For our purposes here we show Characterisation classes only down to the level of the *Support* class (level 3).

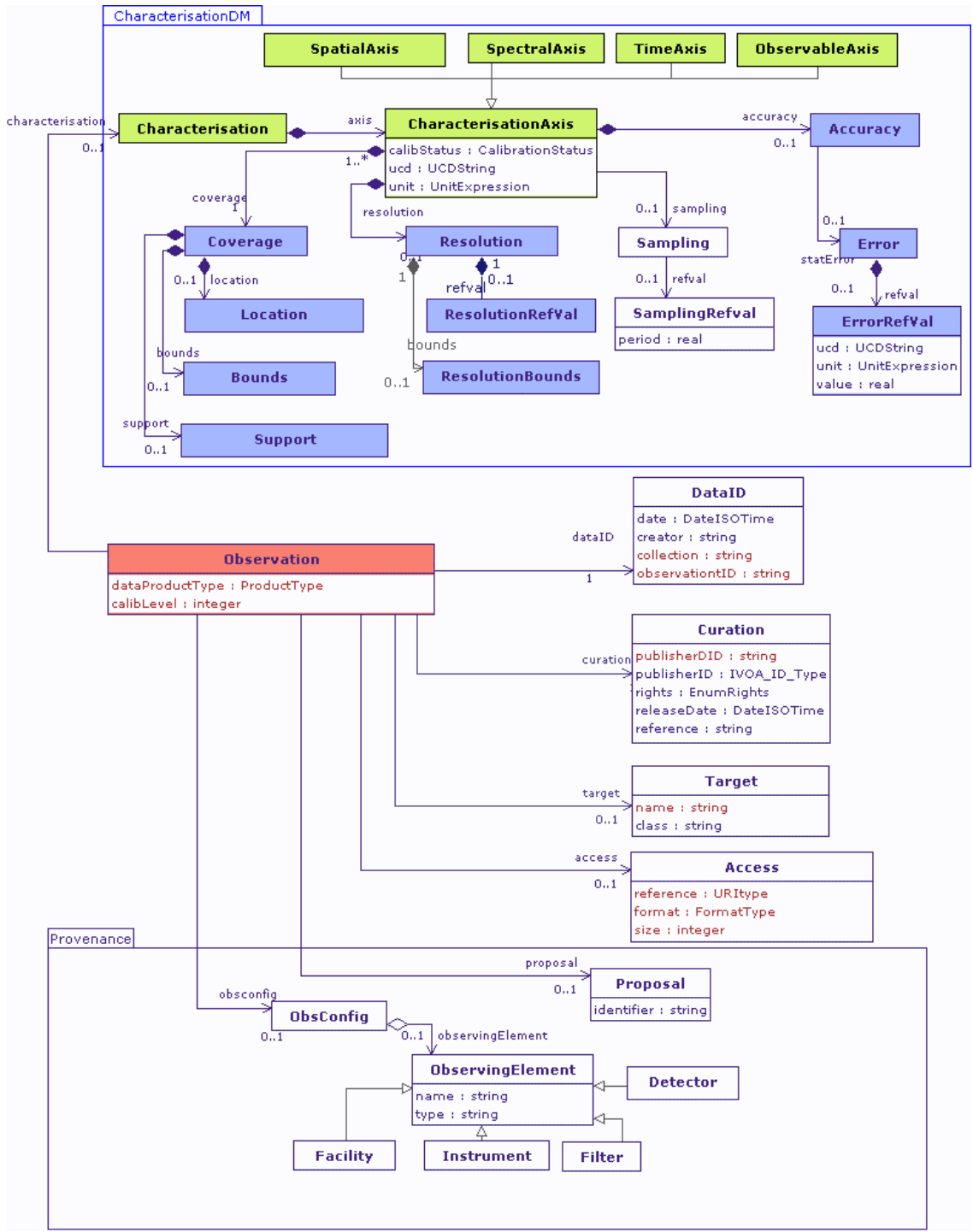

**Figure 2**. Depicted here are the classes used to organize observational metadata. Classes may be linked either via association or aggregation. The minimal set of necessary attributes for data discovery is shown in brown.

For the sake of clarity, the *SpatialAxis*, *SpectralAxis* and *TimeAxis* classes on the diagram are not expanded on the main class diagram. Details for these axes are shown in Figure **3** for the spatial axis, Figure **4** for the spectral axis and Figure **5** for the time axis.

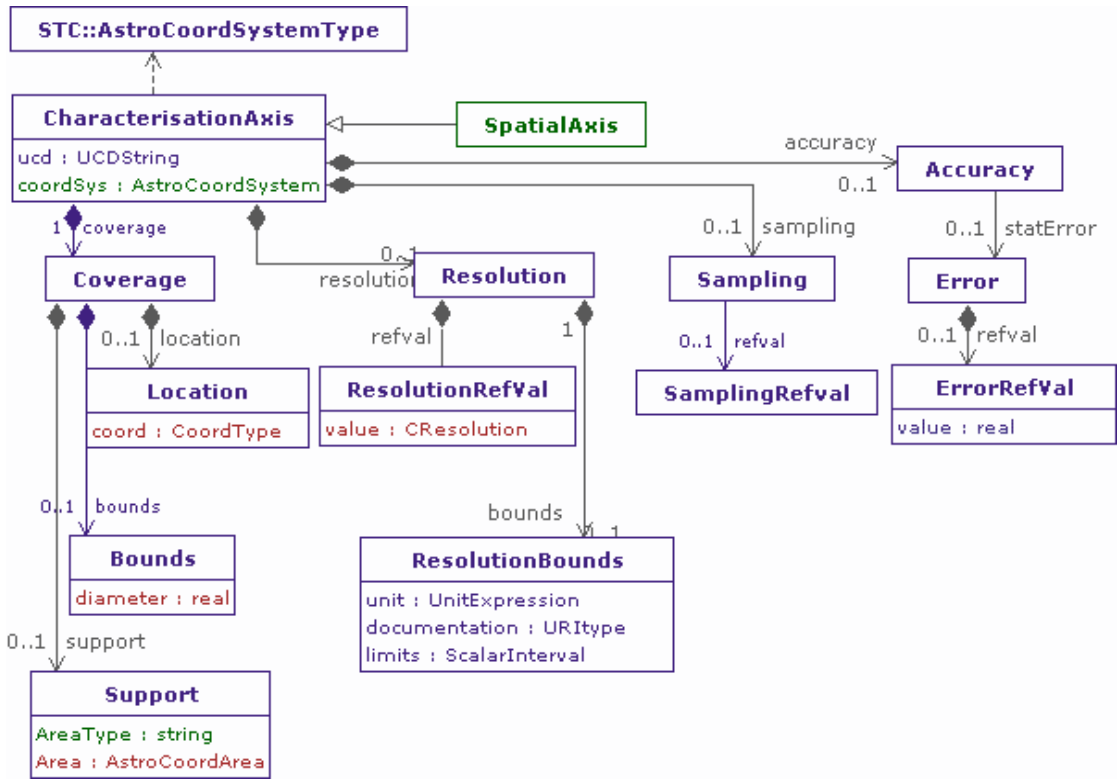

**Figure 3**. Details of the classes linked to the description of the spatial axis for an Observation. All axes in this model inherit the main structure from the *CharacterisationAxis* class, but some peculiar attributes are necessary for Space coordinates.

Details on the ObsCoreDM axes definitions are available in the Characterisation data model standard document (Louys & DataModel-WG., 2008). The hypertext documentation of the model is available (a preliminary version) in the IVOA site under the ObsCore wiki page (http://www.ivoa.net/internal/IVOA/ObsDMCoreComponents/Obscore092011.zip).

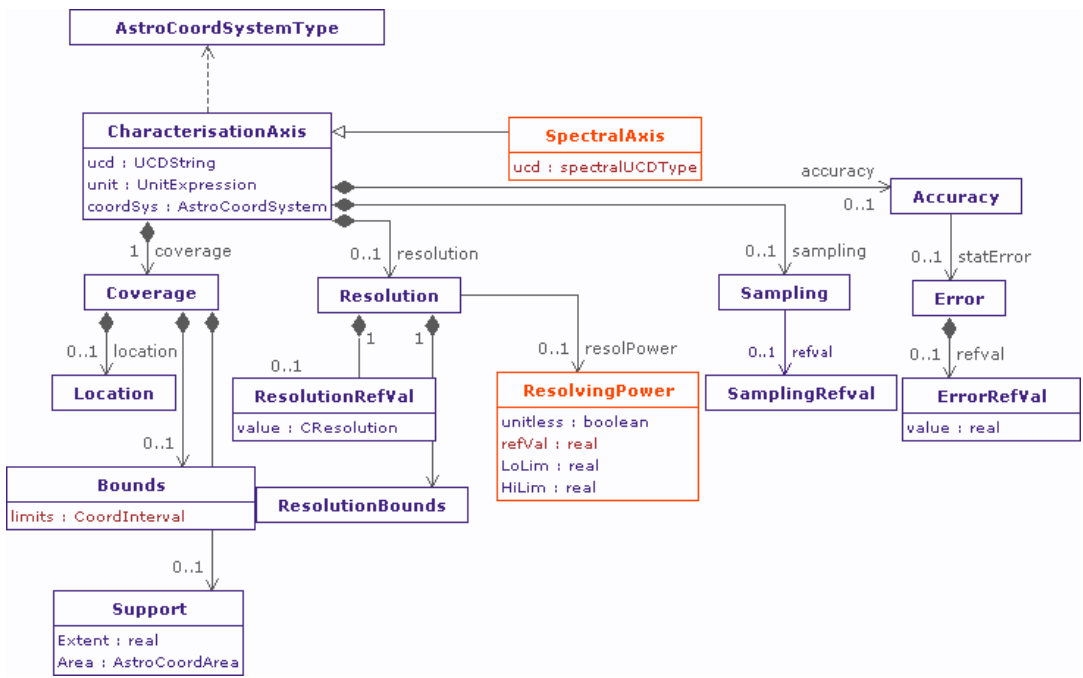

**Figure 4**. Spectral axis: details of the classes necessary to describe the spectral properties of an Observation. UCD and units are essential to disentangle various possible spectral quantities.

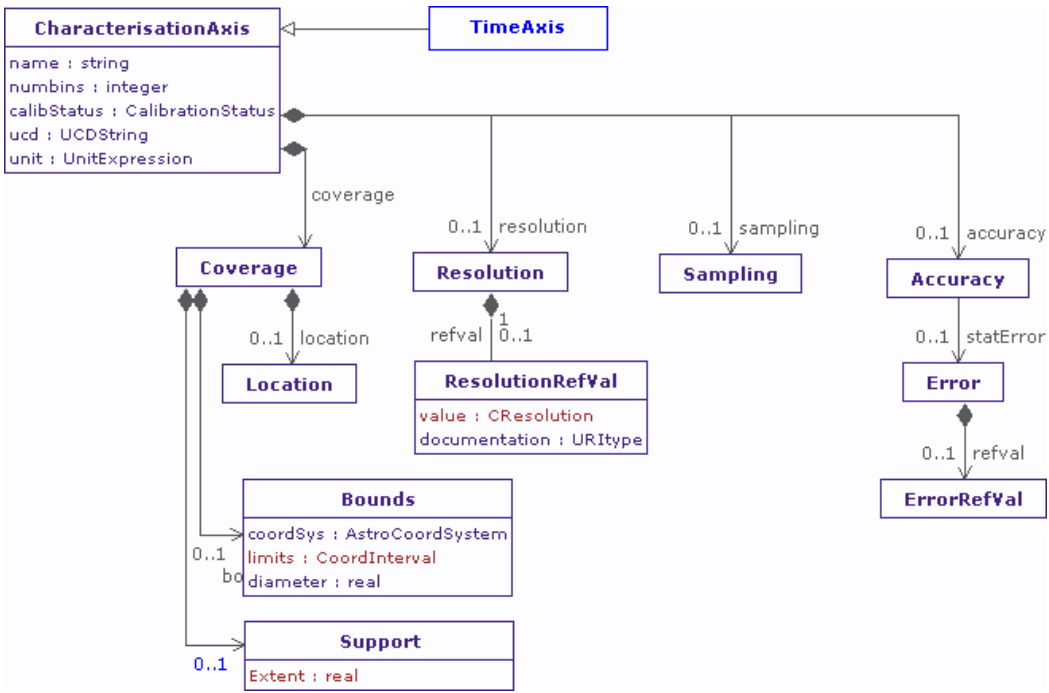

**Figure 5**. The classes from the Characterisation DM used to describe time metadata.

## 3.2. *Main Concepts of the ObsCore Data Model*

The ObsCore data model is the result of the analysis of the data discovery use cases introduced in Chapter 2. Two sets of elements have been identified: those necessary to support the provided use cases, and others that are generally useful to describe the data but are not immediately required to support the use cases. In this section only the first set is described. That set coincides with the set of parameters that any ObsTAP service **must** support. Please refer to appendix B for the detailed description of all obscore elements.

Table 1 lists the data model elements that any ObsTAP implementation **must** support (i.e. a column with such name must exist, though, in some cases, it could be nillable). Provision of these mandatory fields ensures that any query based on these parameters is guaranteed to be understood by all ObsTAP services.

NB: Data model fields are listed here with their **TAP column name** rather than the IVOA data model element identifiers (Utype) to ease readability. See the associated Utypes in **Appendix C**.

| *Column Name* | *Unit* | *Type* | *Description* |
|---|---|---|---|
| dataproduct_type | unitless | string | Logical data product type (image etc.) |
| calib_level | unitless | enum integer | Calibration level {0, 1, 2, 3} |
| obs_collection | unitless | string | Name of the data collection |
| obs_id | unitless | string | Observation ID |
| obs_publisher_did | unitless | string | Dataset identifier given by the publisher |
| access_url | unitless | string | URL used to access (download) dataset |
| access_format | unitless | string | File content format (see in App. BB.5.2 ) |
| access_estsize | kbyte | integer | Estimated size of dataset in kilo bytes |

| target_name | unitless | string | Astronomical object observed, if any |
|---|---|---|---|
| s_ra | deg | double | Central right ascension, ICRS |
| s_dec | deg | double | Central declination, ICRS |
| s_fov | deg | double | Diameter (bounds) of the covered region |
| s_region | unitless | AstroCoordArea | Region covered as specified in STC or ADQL |
| s_resolution | arcsec | float | Spatial resolution of data as FWHM |
| t_min | d | double | Start time in MJD |
| t_max | d | double | Stop time in MJD |
| t_exptime | s | float | Total exposure time |
| t_resolution | s | float | Temporal resolution FWHM |
| em_min | m | double | Start in spectral coordinates |
| em_max | m | double | Stop in spectral coordinates |
| em_res_power | unitless | double | Spectral resolving power |
| o_ucd | unitless | string | UCD of observable (e.g. phot.flux.density) |
| pol_states | unitless | string | List of polarization states or NULL if not applicable |
| facility_name | unitless | string | Name of the facility used for this observation |
| instrument_name | unitless | string | Name of the instrument used for this observation |

**Table 1**. Mandatory fields of the Observation core components data model.

## 3.3.    *Specific Data Model Elements*

In order to support the global data discoverability and accessibility requirements, some new concepts previously not covered by any other data model have to be introduced. This section describes those, which are: the data product type, a classification of the various levels of calibration and processing applied to the data, the file content and format enriched and extended from the concept described in the SSA protocol (Tody & Dolensky, 2008). In addition, a clarification of how the terms *Observation* and *Data Product* are used in the ObsTAP context is provided.

### 3.3.1.    Data Product Type

The model defines a *data product type* attribute to describe the high level scientific classification of the data product being considered. This is coded as a string that conveys a general idea of the content and organization of a dataset. We consider a coarse classification of the types of dataset interesting for science usage, covering: image, cube, spectrum, SED, time series, visibility data, and event data.

- **image** An astronomical image, typically a 2D image with two spatial axes, e.g., a FITS image. The image content may be complex, e.g., an objective-grism observation would be considered a type of image, even though an extracted spectrum would be a Spectrum data product.

- **cube** A multidimensional astronomical image with 3 or more image axes, e.g., a spectral image cube, a polarization cube, a full Stokes radio data cube, a time image cube, etc. The most common format for astronomical "cube" data products is a multidimensional FITS image, however other formats are allowed so long as they are adequately described.

- **spectrum** Any dataset for which spectral coverage is the primary attribute, e.g., a 1D spectrum or a long slit spectrum.

- **sed** A spectral energy distribution, an advanced data product often produced by combining data from multiple observations.

- **timeseries** A one dimensional array presenting some quantity as a function of time. A light curve is a typical example of a time series dataset.

- **visibility** A visibility (radio) dataset of some sort. Typically this is instrumental data, i.e., "visibility data". A visibility dataset is often a complex object containing multiple files or other substructures. A visibility dataset may contain data with spatial, spectral, time, and polarization information for each measured visibility, hence can be used to produce higher level data products such as image, spectra, timeseries, and so forth.

- **event** An event-counting (e.g. X-ray or other high energy) dataset of some sort. Typically this is instrumental data, i.e., "event data". An event dataset is often a complex object containing multiple files or other substructures. An event dataset may contain data with spatial, spectral, and time information for each measured event, although the spectral resolution (energy) is sometimes limited. Event data may be used to produce higher level data products such as images or spectra.

Classification of astronomical data by data product type is inherently ambiguous hence the classification scheme defined here is intentionally kept as simple as possible. The data provider should pick the primary category most appropriate for their data. Values must be specified in lowercase (in order to simplify queries). One of the defined *dataproduct_type* values **must** be used if appropriate for the data product in question, otherwise a NULL value is permitted and a more precise definition of the data product type should be given in *dataproduct_subtype*. Combination of data product types is not allowed, i.e., either one of the above values or NULL must be specified.

Further information on the specific content and format of a data product can be provided by the *dataproduct_subtype* data model field defined in the data model appendix B.1.2 , and by the related *obs_title* (B.3.3) and *access_format* attributes (section 4.7). The intent of *dataproduct_type* is to provide only a general indication of the category to which the data product belongs to facilitate global data discovery.

### 3.3.2. Calibration level

The calibration level concept conveys to the user information on how much data reduction/processing has been applied to the data. It is up to the data providers to consider how to map their own internal classification to the suggested scale here.

**Level 0:** Raw instrumental data, in a proprietary or internal data-provider defined format, that needs instrument specific tools to be handled.

**Level 1:** Instrumental data in a standard format (FITS, VOTable, SDFITS, ASDM, etc.) which could be manipulated with standard astronomical packages.

**Level 2:** Calibrated, science ready data with the instrument signature removed.

**Level 3**: Enhanced data products like mosaics, resampled or drizzled images, or heavily processed survey fields. Level 3 data products may represent the combination of data from multiple primary observations.

The examples in the following section should help illustrate use of the *calib_level* attribute. It is left to the data provider to decide for ambiguous cases.

### 3.3.2.1. Examples of datasets and their calibration level

Here are examples of various datasets, classified according to scheme defined above.

| Data product type | Data collection | Calibration Level | Comments |
|---|---|---|---|
| image | IRAS/NASA | 2 | Science ready data |
| image | IRIS/IRSA | 3 | Recalibrated from infrared IRAS images with removal of the sensor memory effect. |
| image | HDFS/ACS GOODS data | 3 | Image associations mosaicking/stacking |
| spectrum | XMM-Newton EPIC spectra | 1 | Raw instrumental spectrum. |
| cube | EVLA spectral data cube | 2 | Radio spectral data cube in FITS format |
| sed | NED SED | 3 | NED spectral energy distribution |
| event | ROSAT/HEASARC | 1 | Instrumental data |
| visibility | ALMA, Merlin, etc. | 1 | Instrumental data |

**Table 2**. Examples of datasets with calibration level.

### 3.3.3. Observation

ObsTAP and the Observation data model describe *observations in a broad sense* (exactly what comprises an "observation" is not well defined within astronomy and is left up to the data provider to define for their data). ObsTAP also describes archive *data products* (e.g., actual archive files). In general *an "observation" may be composed of multiple individual data products*. In this case all the data products comprising an observation should share the same observation identifier (*obs_id*). The form of the *obs_id* string is up to the data provider so long as it uniquely identifies an observation within the archive. The individual data products comprising an observation may have different data product types, calibration levels, and so forth. ObsTAP only directly supports the description of science data products, i.e., data products which contain science data having some physical (spatial, spectral, temporal) coverage.

In general for instrumental data there are two different approaches for exposing the data from an observation. One can either expose the individual science data products comprising the observation, all sharing the same *obs_id*, or one can expose the entire observation as a single complex instrumental data product. Combinations of the two approaches are also possible.

If the data products comprising an observation are exposed individually then attributes such as the calibration level can vary for different data products, e.g., the raw instrumental data as observed might be level 1, a standard pipeline data product might be level 2, and a custom user-processed data product subsequently published back to the archive might be level 3. All such data products would share the same *obs_id*.

If on the other hand all data from an observation is exposed as a single data product via ObsTAP this will likely be an aggregate of some sort (tar file, directory, etc.) containing multiple files. This latter approach is limited to instrumental data (level 0 or 1), even if objects within the aggregate observation file are higher level. From the perspective of ObsTAP this would be instrumental data, and it is up to the user or client application consuming the data to interpret the meaning of the data elements within the observation.

Which approach is best depends upon the anticipated scientific usage and is up to the data provider to determine. For example if the observational data provided is most commonly used for multi-wavelength analysis, exposing individual high level data products is likely to be the best approach. If the anticipated usage is dominated by complex analysis of instrumental data, then exposing the entire observation as a standard package of instrumental data may be the best approach.

### 3.3.4. File Content and Format

While *dataproduct_type* specifies at a high level what a specific data product is, the *access_format* attribute specifies what is actually in the file. For example, an "image" could be a FITS image, an image embedded in a FITS multi-extension format (MEF) file, a JPEG, etc. A "spectrum" could be represented in the VO-compliant Spectrum format, or in some instrument-specific FITS binary table format. A visibility dataset could be in FITS or ASDM format, or a variety of other radio data formats. A ROSAT or Chandra observation might be presented as a 'tar' file or directory containing instrument-specific observational files. There are many such examples; we give only a few here to illustrate the concept.

Specifying the content and format of a data product is important as special software may be required to do anything useful with the data. The user needs to know exactly what the data product is before deciding to download it for analysis.

See section 4.7 for more details and implementation requirements.

# 4. Implementation of ObsCore in a TAP Service

The ObsCore model must be implemented within Table Access Protocol (TAP) services such that all valid queries can be executed unchanged on any service that implements the model. Additional optional or provider-defined columns are permitted (4.20) so long as all mandatory columns are provided. The protocol does not specify any specific ordering of fields in the query response so long as the mandatory parameters are present in the output stream.

Here we specify an explicit mapping of the model to relational database tables; in the context of TAP this means we are specifying the logical tables as described in the `TAP_SCHEMA` (the TAP-required database schema where the tables and columns exposed by the service are described). This does not necessarily imply that the underlying database will have the identical structure (what is exposed through TAP could be, for example, a database view of the underlying database tables), but in most cases the relationship between `TAP_SCHEMA` description and the underlying tables is straightforward.

| schema_name | table_name | *Description* |
|---|---|---|
| ivoa | ivoa.ObsCore | ObsCore 1.0 |

**Table 3**. `TAP_SCHEMA.tables` for implementation of the ObsCore model.

| table_name | column_name | *data type* | *units* | *constraint* |
|---|---|---|---|---|
| ivoa.ObsCore | dataproduct_type | adql:VARCHAR | | |
| ivoa.ObsCore | calib_level | adql:INTEGER | | not null |
| ivoa.ObsCore | obs_collection | adql:VARCHAR | | not null |
| ivoa.ObsCore | obs_id | adql:VARCHAR | | not null |
| ivoa.ObsCore | obs_publisher_did | adql:VARCHAR | | not null |
| ivoa.ObsCore | access_url | adql:CLOB | | |
| ivoa.ObsCore | access_format | adql:VARCHAR | | |

| ivoa.ObsCore | access_estsize | adql:INTEGER | kbyte | |
| ivoa.ObsCore | target_name | adql:VARCHAR | | |
| ivoa.ObsCore | s_ra | adql:DOUBLE | deg | |
| ivoa.ObsCore | s_dec | adql:DOUBLE | deg | |
| ivoa.ObsCore | s_fov | adql:DOUBLE | deg | |
| ivoa.ObsCore | s_region | adql:REGION | deg | |
| ivoa.ObsCore | s_resolution | adql:DOUBLE | arcsec | |
| ivoa.ObsCore | t_min | adql:DOUBLE | d | |
| ivoa.ObsCore | t_max | adql:DOUBLE | d | |
| ivoa.ObsCore | t_exptime | adql:DOUBLE | s | |
| ivoa.ObsCore | t_resolution | adql:DOUBLE | s | |
| ivoa.ObsCore | em_min | adql:DOUBLE | m | |
| ivoa.ObsCore | em_max | adql:DOUBLE | m | |
| ivoa.ObsCore | em_res_power | adql:DOUBLE | | |
| ivoa.ObsCore | o_ucd | adql:VARCHAR | | |
| ivoa.ObsCore | pol_states | adql:VARCHAR | | |
| ivoa.ObsCore | facility_name | adql:VARCHAR | | |
| ivoa.ObsCore | instrument_name | adql:VARCHAR | | |

**Table 4**. List of the minimal set of data model fields to implement for an ObsTAP service. See tables on page 50 in Appendix C for the full description of the TAP_SCHEMA.columns table.

Table 3 and Table 4 provide the primary information needed to describe the ObsCore model in terms of `TAP_SCHEMA` tables and columns. The "constraint" specified in Table 4 above is not part of the `TAP_SCHEMA.columns` description, but is required by the ObsCore model and specified here to make this clear to implementers. Additional standard content for the individual columns is specified below.

## 4.1.　Data Product Type (dataproduct_type)

The *dataproduct_type* column contains a simple string value describing the primary nature of the data product. It should assume one of these string values: **image**, **cube**, **spectrum**, **sed**, **timeseries**, **visibility**, or **event**. These values are described in section 3.3.1. A NULL value is permitted, but only in the event that none of the proposed values can be used to describe the dataset. The optional field *dataproduct_subtype* (B.1.2) may be used to more precisely define the nature of the dataset. Values in the *dataproduct_typ*e column **must** be written in lower case. Specifying this field along with the desired spatial and spectral coverage will be enough to discover data of interest in many common cases.

Usage: select * from ivoa.ObsCore where dataproduct_type='image' returns only image data.

## 4.2.　Calibration Level (calib_level)

The *calib_level* column tells the user the amount of calibration processing that has been applied to create the data product. *calib_level* **must** assume one of the following integer values: 0 (instrumental or raw data in a non-standard/proprietary format), 1 (instrumental data in a standard format, e.g. FITS), 2 (calibrated data in standard format, with instrument signature removed), and 3 (more highly processed data product). Please refer to section 3.3.2 for a full description. Data providers decide which value best describes their data products.

Values in the *calib_level* column **must not** be NULL.

Usage: `select * from ivoa.ObsCore where calib_level >2` returns enhanced data products.

## 4.3. Collection Name (obs_collection)

The *obs_collection* column identifies the data collection to which the data product belongs. A *data collection* is any logical collection of datasets which are alike in some fashion. Typical data collections might be all the data from a particular telescope, instrument, or survey. The value is either the registered shortname for the data collection, the full registered IVOA identifier for the collection, or a data provider defined shortname for the collection. Often the collection name will be set to the name of the instrument that generated the data. In that case we suggest specifying the collection name as a string composed of the facility name, followed by a slash, followed by the instrument name.

Examples : HST/WFPC2, VLT/FORS2, CHANDRA/ACIS-S.

There are other cases where it makes no sense to use the instrument name, may be because the data product used data from different instruments or facilities, or for other reasons. Examples: SDSS-DR7, etc.

In practice this is not a very precisely defined field. What is important is for the data provider to use a collection name which is familiar to astronomers and discriminative to point easily on datasets of interest.

Values in the *obs_collection* column **must not** be NULL.

## 4.4. Observation Identifier (obs_id)

The *obs_id* column defines a unique identifier for an observation. In the case where multiple data products are available for an observation (e.g. with different calibration levels), the *obs_id* value will be the same for each data product comprising the observation. This is equivalent to the dataset name for many archives where dataset name could have many files associated with them. The returned *obs_id* for an archival observation should remain identical through time for future reference.

In the case of some advanced data products (with calib_level =3), the data product may be the result of combining data from multiple primary (physical) observations. In this case the resulting data product is a new processed "observation" to which a new unique observation identifier should be assigned. If the advanced processing results in several associated data products they should share the same *obs_id*. Describing the provenance of such an advanced data product is possible, but is out of scope for ObsTAP.

Values in the *obs_id* column **must not** be NULL.

## 4.5. Publisher Dataset Identifier (obs_publisher_did)

The *obs_publisher_did* column contains the IVOA dataset identifier (Plante & al., 2007) for the published data product. **This value must be unique within the namespace controlled by the dataset publisher** (data center). The value will also be globally unique since each publisher has a unique IVOA registered publisher ID. The same dataset may however have more than one publisher dataset identifier if it is published in more than one location; the creator DID, if defined for the given dataset, would be the same regardless of where the data is published.

The returned *obs_publisher_did* for a static data product should remain identical through time for future reference.

Values in the *obs_publisher_did* column **must not** be NULL.

## 4.6. Access URL (access_url)

The *access_url* column contains a URL that can be used to download the data product (as a file of some sort).

We specify the data type as CLOB (character large object) in the TAP service so that users will know they can only use the *access_url* column in the SELECT clause of a query. That is, users cannot specify this column as part of a condition in the WHERE clause and implementers are free to generate the URL on the fly during output (rather than being forced to store it statically in the database).

More details are given on the use of CLOB data types for the TAP SCHEMA in the TAP Standard document (Dowler, Tody, & Rixon, 2010), section 2.5 Table upload.

Access URLs are not guaranteed to remain valid and unchanged indefinitely. To access a specific data product after a period of time (e.g., days or weeks) a query should be performed (e.g., using *obs_publisher_did*) to obtain a fresh access URL.

## 4.7. Access Format (access_format)

The *access_format* column specifies the format of the data product if downloaded as a file. This data model field is important both for data discovery and for the client to evaluate whether it will be able to actually use the data product if downloaded.

MIME types are often used to specify file formats in existing protocols such as HTTP (Authority Internet Assigned Numbers, 2007). However when dealing with astronomical observations as in ObsTAP services, more information about the format of the data is required than can be specified by conventional MIME types. For instance we might want to distinguish between various formats like multi-extension FITS (e.g. for CCD mosaic instruments or MUSE IFU data), or ASDM (e.g. for ALMA or other interferometric observations). Even for something as fundamental to astronomy as FITS binary table there is currently no standardized MIME type other than the generic `application/FITS`.

While standard MIME types are limited when it comes to describing the many data formats actually in use within astronomy, they are ideal for specifying common file types such as HTML and XML, the various graphics file types, text, PDF, and so forth, all of which can be used to describe aspects of observational data. Furthermore the MIME type scheme is extensible, allowing new formats which are not yet standardized to be specified. Hence what we propose here is to adopt the MIME type mechanism to describe the *file format* of a science data product, defining new custom types as needed. Note this is distinct from the *science content* which is specified by the data product type and subtype. The same content can potentially be represented in multiple formats hence these are distinct.

The following table illustrates some common astronomical file formats. This list is by no means intended to be comprehensive; rather it illustrates the approach while defining standard values for some common formats. Some randomly selected formats are included to illustrate the approach. We can extend this list as experience is gained using ObsTAP to describe actual data archives.

| MIME-type | Shortname | Definition |
|---|---|---|
| image/fits | fits | Any multidimensional regularly sampled FITS image or cube |
| image/jpeg | jpeg | A 2D JPEG graphic image (likewise for GIF, PNG, etc.) |
| application/fits | fits | Any generic FITS file |
| application/x-fits-bintable | bintable | A FITS binary table (single BINTABLE extension) |
| application/x-fits-mef | mef | A FITS multi-extension file (multiple extensions) |

| application/x-fits-uvfits | uvfits | A FITS file in UVFITS format (likewise SDFITS etc.) |
|---|---|---|
| application/x-fits-euro3d | euro3d | A FITS file in Euro3D format (multiobject spectroscopy) |
| application/x-votable+xml | votable | Any generic VOTable file |
| application/x-asdm | asdm | ALMA science data model (final export format still TBD) |
| application/pdf | pdf | Any PDF file |
| text/html | html | Text in HTML format |
| text/xml | xml | Any generic XML file |
| text/plain | txt | Any generic text file |
| text/csv | csv | Tabular data in comma separated values format |
| text/tab-separated-values | tsv | Tabular data in tab separated values format |
| application/x-tar | tar | Multiple files archive in TAR format |
| application/zip | zip | Multiple files archive in ZIP format |
| application/x-directory | dir | Multiple files archive returned as a text list |
| image/x-fits-gzip | fits | A GZIP-compressed FITS image |
| image/x-fits-hcompress | fits | A FITS image using HCOMPRESS compression |
| application/x-tar-gzip | gtar | A GZIP-compressed TAR file (x-gtar also sometimes used) |

The value of *access_format* should be a MIME type, either a standard MIME type, an extended MIME type from the above table, or a new custom MIME type defined by the data provider. The short names suggested here are not used directly by *access_format*.

Custom file formats should be specified using a MIME type such as "*application/x-<whatever>*". This can be used for any file format including custom binary file formats.

Observational datasets consisting of multiple instrument-specific files may be exposed in formats like application/x-directory, application/x-tar or application/x-tar-gzip. Details of the package content and how to access inner data products will be described in a separate effort, called "Data Linking", currently in development in the DM and DAL Working group. See the example presented in section C.1.1 .

Compression is inherent in some file formats, e.g., ZIP or JPEG. In other formats it is optional and is indicated by having multiple versions of the format, e.g. image/fits or image/x-fits-gzip for a GZIP-compressed FITS image (the "x-" prefix is required for anything which is not a registered standard MIME type).

## 4.8. Estimated Download Size (access_estsize)

The *access_estsize* column contains the approximate size (in kilobytes) of the file available via the *access_url*. This is used only to gain some idea of the size of a data product before downloading it, hence only an approximate value is required. Provision of dataset size estimates is important whenever it is possible that datasets can be very large.

## 4.9. Target Name (target_name)

The *target_name* column contains the name of the target of the observation, if any. This is typically the name of an astronomical object, but could be the name of a survey field.

The target name is most useful for output, to identify the target of an observation to the user. In queries it is generally better to refer to astronomical objects by position, using a name resolver to convert the target name into a coordinate (when possible).

## 4.10. Central Coordinates (s_ra, s_dec)

The coordinate system in which coordinates are expressed is ICRS. The *s_ra* column specifies the ICRS Right Ascension of the center of the observation. The *s_dec* column specifies the ICRS Declination of the center of the observation.

## 4.11. Spatial Extent (s_fov)

The *s_fov* column (1D size of the field of view) contains the approximate size of the region covered by the data product. For a circular region, this is the diameter (not the radius). For most data products the value given should be large enough to include the entire area of the observation; coverage within the bounded region need not be complete, for example if the specified FOV encompasses a rotated rectangular region. For observations which do not have a well-defined boundary, e.g. radio or high energy observations, a characteristic value should be given.

The *s_fov* attribute provides a simple way to characterize and use (e.g. for discovery computations) the approximate spatial coverage of a data product. The spatial coverage of a data product can be more precisely specified using the *s_region* attribute (4.12).

## 4.12. Spatial Coverage (s_region)

The *s_region* column can be used to precisely specify the covered spatial region of a data product.

It is often an exact, or almost exact, representation of the illumination region of a given observation defined in a standard way by the concept of Support in the Characterisation data model.

We specify the data type as the logical type `adql:REGION` so that users can specify spatial queries using a single column and in a limited number of ways. If included in the select list of the query, the output is always an STC-S string as described in (Dowler, Tody, & Rixon, 2010) [section 6]. In the WHERE clause, the *s_region* column can be used with the ADQL geometry functions (INTERSECTS, CONTAINS) to specify conditions; the service will generally have to translate these into native SQL that enforces the same conditions or a suitable approximation. Implementers may approximate the spatial query conditions by translating the INTERSECTS and CONTAINS function calls in the query.

In addition, ADQL specifies several functions, which may take the *s_region* column as an argument: AREA, CENTROID, and COORDSYS. The AREA function returns the area (in sq. deg.) of the observed region. In cases where the *s_region* itself is an approximation (a bounding box, for example), this function should still return the actual value. This may be implemented by computing and storing the area in a separate column and converting the AREA (s_region) function call into a column reference in the query. The CENTROID function returns an ADQL POINT value; if used in the select list the output is always an STC-S string as described in (Dowler, Tody, & Rixon, 2010) [section 6]. The coordinates must be the same as those found in the *s_ra* and *s_dec* columns, which are provided for convenience. The COORDSYS function returns the coordinate system used for the *s_region*; in the ObsCore model implementation here this is restricted to ICRS, so this can be implemented by converting the COORDSYS (s_region) function call to a constant in the query.

Region computations are an advanced query capability which may not be supported by all services. Services should however specify *s_region* when possible to more precisely specify the spatial coverage of an observation.

## 4.13. Spatial Resolution (s_resolution)

The *s_resolution* column specifies a reference value chosen by the data provider for the estimated spatial resolution of the data product in arcseconds. This refers to the smallest spatial feature in the observed signal that can be resolved.

In cases where the spatial resolution varies across the field the best spatial resolution (smallest resolvable spatial feature) should be specified. In cases where the spatial frequency sampling of an observation is complex (e.g., interferometry) a typical value for spatial resolution estimate should be given; additional characterization may be necessary to fully specify the spatial characteristics of the data.

## 4.14. Time Bounds (t_min, t_max)

The *t_min* column contains the start time of the observation specified in MJD. The *t_max* column contains the stop time of the observation specified in MJD. In case of data products result of the combination of multiple frames, *t_min* must be the minimum of the start times, and *t_max* as the maximum of the stop times.

## 4.15. Exposure Time (t_exptime)

The *t_exptime* column contains the exposure time. For simple exposures, this is just *t_max* - *t_min* expressed in seconds. For data where the detector is not active at all times, (e.g. data products made by combining exposures taken at different times), the *t_exptime* will be smaller than *t_max* - *t_min*. For data where the *t_exptime* is not constant over the entire data product, the median exposure time per pixel is a good way to characterize the typical value. In some cases, *t_exptime* is generally used as an indicator of the relative sensitivity (depth) *within a single data collection* (e.g. *obs_collection*); data providers should supply a suitable relative value when it is not feasible to define or compute the true exposure time.

In case of targeted observations, on the contrary the exposure time is often adjusted to achieve similar signal to noise ratio for different targets.

## 4.16. Time Resolution (t_resolution)

The *t_resolution* column is the minimal interpretable interval between two points along the time axis. This can be an average or representative value. For products with no sampling along the time axis, the *t_resolution* could be set to the exposure time or could be null. That way one could compose a WHERE clause like: `WHERE t_resolution < t_exptime` to find those products which are time resolved.

This implementation preference avoids dealing with undefined data model fields as originally considered in the Characterisation data model for unresolved time axis.

## 4.17. Spectral Bounds (em_min, em_max)

The *em_min* column specifies the minimum spectral value observed, expressed as a vacuum wavelength in meters.

The *em_max* column contains the maximum spectral value observed, expressed as a vacuum wavelength in meters.

As mentioned in the data model in Appendix B, at least 3 physical quantities could in principle be used to represent the spectral axis: energy, wavelength or frequency; which is most appropriate depends upon the observation domain. For ObsTAP we are less concerned with how to present data to the user than with providing a simple and uniform way to describe astronomical data, hence we restrict the spectral bounds units to wavelength in meters in vacuum. Conversion to other quantities could be performed either on the client side for an application encapsulating queries, and/or on the server side, for a data provider to expose its data from other regimes to ObsTAP queries.

## 4.18. Spectral Resolving Power (em_res_power)

The *em_res_power* column contains the typical or characteristic spectral resolving power defined as $\lambda/\delta\lambda$. The value is dimensionless.

## 4.19. Observable Axis Description (o_ucd)

The *o_ucd* column specifies a UCD (Preite Martinez, Derriere, Delmotte, Gray, & al., 2007) describing the nature of the observable within the data product. The observable is the measured quantity, for example photon counts or flux density stored in the pixel value within an image. Often for optical astronomical images the value would be `phot.count`; for fully flux calibrated data a value such as `phot.flux.density` (usually specified in Jy) would be used. Any valid UCD is permitted. If no appropriate UCD is defined the field should be left NULL (the IVOA provides a process by which new UCDs can be defined).

## 4.20. Additional Columns

Service providers may include additional columns in the *ivoa.ObsCore* table to expose additional metadata. These columns must be described in the *TAP_SCHEMA.columns* table and in the output from the VOSI-tables resource ([VOSI] Grid and Web service WG, 2010). Users may access these columns by examining the column metadata for individual services and then using them explicitly in queries or by selecting all columns in the query (e.g. "select * from ivoa.ObsCore ..." in an ADQL query). In order to provide homogeneity in the keywords used as optional fields, we recommend where possible to use the items defined in the full data model (Appendix B) and flagged as optional. ObsTAP compliant services will support all columns defined as mandatory and possibly some of the optional ones. Queries built up using additional columns defined specifically for a given archive might not be portable.

# 5. Registering an ObsTAP Service

The standard identifier for the ObsCore model described here is `ivo://ivoa.net/std/ObsCore/v1.0`.

The ObsCore data model will be registered using this identifier and the StandardsRegExt (standards registry extension).

TAP services that implement the ObsCore model should be registered to indicate this fact so that users can easily find all services that accept ObsCore queries. This can be done in any registry by using the keyword "ObsCore" to describe the service. In addition, fine-grained registries may include the complete VODataService table set description.

The TAPRegExt[1] (Table Access Protocol registry extension) (Demleitner, Plante, Dowler, Rixon, & Taylor, 2011) provides a mechanism (the '**dataModel**' element) to list one or more data models that are supported by a TAP service. The data model support uses the ivo standard identifier (above)[2]. One or more '**dataModel**' elements may be included as child elements of the '**capability**' element describing the TAP service (the '**capability**' element) with `standardID="ivo://ivoa.net/std/TAP"`).

In general, the data model support in TAPRegExt can be used when a TAP service contains tables and columns described with Utypes from a standard data model; it is not generally necessary to have all the Utypes (e.g. the complete model).However, since the ObsCore data model is a physical model designed specifically to be implemented in TAP services, the standard identifier must only be used to specify data model support in the TAPRegExt if the ivoa.ObsCore table is available and contains all the mandatory columns [3].

---

[1] The TAPRegExt is an internal working draft as of June 2011; service providers should use it to describe TAP services once completed.

[2] Since the TAPRegExt is a draft and subject to revision, we cannot provide a concrete example at this time.

# 6. Implementation Examples

ObsTAP implementations will be described in a separate IVOA Note.

Examples of supported use-cases are also provided at the following URL:
http://www.cadc.hia.nrc.gc.ca/cvo/ObsCore

# 7. Changes from Earlier Versions

Version 0.2 to 1.0 Spring 2010:

- Include implementation part in section 3
- Fix underscore character in most places
- Include data model summary table for all fields in appendix A
- Add a status column for each field M or R or O
- Update tables em_domain moved up (minimal change)

Version 1.0 May 2010:

- Section 5: re-write XMM SSC ObsTAP service description
- Introduce use case shortly at beginning and point to appendix
- Moved data model summary table back to data model section

Version 1.0 Dec 2010 to February 2011:

- Converted document from Latex to Word
- Revised data model column names to make more consistent
- Moved table showing full data model to an appendix
- Added a short table of only the mandatory fields to the DM intro
- Many small edits to make text more readable and correct
- Edit Use-case part and provide query examples
- Added references and citations
- Added figure for time Axis
- Changed appendix C for a data modeling orientation

Version 1.0-20110227:

---

[3] Additional columns with optional ObsCore Utypes, Utypes from other data models, or no Utypes at all are allowed

- Merged appendix A and B into A
- Flesh out appendix B for a data modeling orientation
- Update Utype syntax in all tables
- Re-organize appendix C and D for TAP_SCHEMA.columns and example of table initialization

Version 1.0-20110415:

- Insert updates after mailing list and wiki discussion.
- Use Camel Case for Utypes in this document

Version 1.0-2011 May

- D.Tody updates and fixes all the formatting of the document (Toc, section numbers, etc.)
- M.Louys updates bibliographic references and cross-referencing between sections within the document

Version 1.0-2011June

- ML updates VOArchitecture figure with TapRegExt box instead of SimpleDALRegExt box
- Tab 6. : Correct Utype using CResolution STC type

Remove ADQL query examples and update implementation webpage to show them in action instead of ADQL text only

Version 1.0-2011September

- Add-ons to cover comments from the implementation feedback at Chandra's data center
- Typical values for spatial resolution and ranges (4.13, B.6.1.3)
- Noise type on the observable axis as o_stat_error_type ( see B.6.7)
- Example of package of heterogeneous data products in C1.1

# Appendix A: Use Cases in detail

The ability to discover data of a certain kind (images, spectra, cubes, etc.) according to scientific criteria (e.g., a given sky position, spectral coverage including spectral line X, spatial resolution better than Y, resolving power greater than Z) is central to archival astronomy. A special Take Up Committee of the IVOA was formed in 2009 to stimulate IVOA work in the area of catalogue-based science data access to allow astronomers to easily query and access scientific data. This committee came up with a list of data discovery use cases expressed as a set of constraints on selected scientific parameters to be used to query for datasets of interest. The full list of use cases is summarized below.

Please note that for most science cases, a full TAP implementation is required for them to work as well as STC regions support. (Rots, 2007)

Some of the use-cases listed by the committee require advanced functionalities like "search by type", "query from an input list", and have not been fully developed here.

Once a full TAP implementation is available, we will expand these science cases into working examples so they could be used as template and/or teaching examples.

## *Simple Examples*

### Simple Query by Position

Show me a list of all data that satisfies:

    I.   Datatype=any
   II.  contains RA=16.0 and DEC=40.0

These data would be searched on all VO services by sending the following query:

```
SELECT * FROM ivoa.Obscore WHERE
CONTAINS(POINT('ICRS',16.0,40.0),s_region)=1
```

This query could be submitted to a remote TAP service using the *curl* application as follows (in this example a CADC TAP service is referenced):

```
curl -v -L -d "REQUEST=doQuery&LANG=ADQL&QUERY=select * from ivoa.ObsCore
where CONTAINS(POINT('ICRS',16.0,40.0 ),s_region)=1"
"http://www.cadc.hia.nrc.gc.ca/caom/sync"
```

More constraints are added in the following use-case (1.3).

### Query by both Spatial and Spectral Attributes

Show me a list of all data that satisfies:

    I.   DataType=Image
   II.  Spatial resolution better than 0.3 arc seconds
  III.  Filter = J or H or K
  IV.  RA between 16 hours and 17 hours
   V.  DEC between 10 degrees and 11 degrees

Such a query needs to compute RA in degrees, extract information from Filter and adjust spectral intervals for search.

```
SELECT * FROM ivoa.Obscore
WHERE dataproduct_type='image'
AND s_resolution < 0.3 AND s_ra > 240 AND s_ra < 255
AND s_dec > 10 AND s_dec < 11
AND (em_min > 2.1e-06 AND em_max < 2.4e-06)
      OR(em_min >= 1.6e-06 AND em_max <= 1.8e-06)
      OR(em_min >= 1.2e-06 AND em_max <= 1.4e-06)
```

A similar query could be submitted to a remote TAP service using the *curl* application as follows (in this example a CADC TAP service is referenced):

```
curl -v -L -d "REQUEST=doQuery&LANG=ADQL&QUERY=
select * from ivoa.ObsCore where dataproduct_type='image' AND s_resolution < .3
AND s_ra >240 AND s_ra < 255 AND s_dec > 10 and s_dec < 11
and (em_min > 2.1e-06 AND em_max < 2.4e-06)
OR(em_min >= 1.6e-06 AND em_max <= 1.8e-06)
OR(em_min >= 1.2e-06 AND em_max <= 1.4e-06)"
"http://www.cadc.hia.nrc.gc.ca/caom/sync"
```

## *A.1    Discovering Images*

### A.1.1.    Use case 1.1

Show me all observations satisfying:

    I.   DataType = any
   II.  Energy includes 5 keV

III.  RA includes 16.00
IV.  DEC includes +10
V.  Exposure time > 10 ks

```
SELECT * FROM ivoa.Obscore
WHERE em_min < 2.48E-10 AND em_max > 2.48 E-10
AND CONTAINS(POINT('ICRS',16.0,10.0),s_region) = 1
AND t_exptime > 10000
```

## A.1.2.  Use case 1.2

Let me input a list of RA and DEC coordinates and show me spatially coincident observations satisfying:

I.  Imaging or spectroscopy data
II.  Includes one or more of the RA,DEC values in the list (LIST=SERVICE REQ)
III.  Includes both a wavelength in the range 5000-9000 angstroms AND an X-ray image (AND=SERVICE REQ)

This use case may need several steps to select images from X-RAY domain, select image and spectra on optical domain and compute the overlap.

It requires two functionalities from the service:

- LIST=SERVICE REQ, to query on lists of positions given as input
- AND=SERVICE REQ, to compute the intersection between two response lists.

## A.1.3.  Use case 1.3

Show me a list of all observations satisfying:

I.  DataType=Image
II.  Spatial resolution better than 0.3 arcseconds
III.  Filter = J or H or K
IV.  RA between 16 hours and 17 hours
V.  DEC between 10 degrees and 11 degrees

```
SELECT * FROM ivoa.Obscore
WHERE dataproduct_type='image'
AND s_resolution < 0.3
AND (
( -- J band approximated
(em_min + em_max)/2 BETWEEN 1.234E-6 - 162E-9 AND 1.234E-6 + 162E-9
AND
(em_max - em_min) BETWEEN 0.5 * 162E-9 AND 1.5 * 162E-9
)
OR
( -- H band approximated
(em_min + em_max)/2 BETWEEN 1.662E-6 - 251E-9 AND 1.662E-6 + 251E-9
AND
(em_max - em_min) BETWEEN 0.5 * 251E-9 AND 1.5 * 251E-9
)
OR
( -- J band approximated
(em_min + em_max)/2 BETWEEN 2.159E-6 - 262E-9 AND 2.159E-6 + 262E-9
AND
(em_max - em_min) BETWEEN 0.5 * 262E-9 AND 1.5 * 262E-9
)
AND s_ra BETWEEN 16*15 AND 17*15
AND s_dec BETWEEN 10 and 11
```

### A.1.4. Use case 1.4

Show me a list of all observations that satisfying:

    I.   DataType=Image
    II.  Wavelength=V or I or Z
    III. Spatial Resolution < 0.7 arcsec FWHM
    IV. Exposure Time > 1000 seconds
    V.  Data Quality: Fully Calibrated

### A.1.5. Use case 1.5

Show me all data that satisfies:

    I.   DataType=IFU
    II.  DataQuality: Fully Calibrated
    III. ObjectClass=quasar (SERVICE REQ + NEEDS ANOTHER SERVICE (CATALOGUE)
    IV. Redshift > 3
    V.  Radio flux > 1 mJy

We assume here that data providers will expose IFU data using dataproduct_type='cube'.

```
SELECT * FROM ivoa.Obscore
WHERE dataproduct_type='cube'
AND calib_level > 1
AND CONTAINS(POINT('ICRS', quasar_ra, quasar_dec), s_region) = 1
```

### A.1.6. Use case 1.6

For an input list of RA, DEC, Modified Julian Date (MJD), show me all data that satisfies (LIST=SERVICE REQ)

    I.   DataType=imaging
    II.  RA,DEC include the value of the list and Observation date is within 1 day of the MJD value

```
SELECT * FROM ivoa.Obscore
WHERE dataproduct_type='image'
AND CONTAINS(POINT('ICRS',user_ra,user_dec), s_region) = 1
AND | (t_max + t_min)/2 - user_date | < 1 d
```

## A.2. Discovering spectral data

### A.2.1. Use case 2.1

Show me a list of all data that satisfies:

    I.   DataType=Spectrum
    II.  Energy spans 1 to 5 keV
    III. Total counts in spectrum > 100
    IV. Exposure time > 10000 seconds
    V.  Data Quality: Fully Calibrated

### A.2.2. Use case 2.2

Show me a list of all data that satisfies:

    I.   DataType=Spectrum
    II.  Wavelength includes 6500 angstroms
    III. Spectral Resolution better than 15 angstroms

IV. Spatial Resolution better than 2 arcseconds FWHM
V. Exposure Time > 3600 seconds
VI. Data Quality = Any

```
SELECT * from ivoa.Obscore
WHERE dataproduct_type='spectrum'
AND em_min < 650E-9
AND em_max > 650E-9
AND em_res_power > 6500/15.
AND s_resolution < 2
AND t_exptime > 3600
```

### A.2.3. Use case 2.3

Show me a list of all data that satisfies:

I. Emission line width Halpha > 2000 km/s FWHM (SERVICE REQ+NEEDS OTHER SERVICE)

## A.3. Discover multi-dimensional observations

### A.3.1. Use case 3.1

Show me a list of data with:

I. DataType=cube (IFU spectroscopy?)
II. RA,DEC includes value RA1,DEC1
III. Field size > 100 square arcseconds
IV. DataSensitivity = deep
V. Spectral resolution better than 5 angstroms FWHM

### A.3.2. Use case 3.2

Show me a list of all data that satisfies:

I. DataType=cube with 3 dimensions
II. Axes includes Velocity
III. Axes includes RA
IV. Axes includes DEC
V. Velocity Resolution better than 50 km/s
VI. RA includes 16.000
VII. Dec includes +41.000

NB: in this case optional data model fields related to redshift axis can be used using redshift_ucd=`spect.DopplerVeloc`, for instance.

### A.3.3. Use case 3.3

Show me a list of all data that satisfies:

I. DataType=cube
II. RA includes 16.00
III. Dec includes +41.00
IV. Wavelength includes 6500 angstroms
V. Wavelength includes 4000 angstroms
VI. Spectral resolution better than 5 angstroms
VII. Exposure time more than 3600 seconds
VIII. Data Quality: Fully Calibrated

### A.3.4. Use case 3.4

Show me a list of all data that satisfies:

       I. DataType=Cube with 3 dimensions
      II. Axes includes FREQ
      III. Axes includes RA
      IV. Axes includes DEC
      V. Velocity Resolution better than 1 km/s
      VI. RA includes 83.835000
      VII. Dec includes -5.014722
      VIII. Rest Frequency = 345.795990 GHz
      IX. VLSRK in the range [6.0, 10.0]

### A.3.5. Use case 3.5

Show me a list of all data that satisfies:

      I. DataType=Cube with 3 dimensions
      II. Axes includes FREQ
      III. Axes includes RA with > 100 pixels
      IV. Axes includes DEC with > 100 pixels
      V. Frequency extent > 500 MHz
      VI. Rest Frequency = 345.795990 GHz appears in band
      VII. The redshift is not specified, but should default to zsource for the target.

      NB: I to V are supported in ObsTAP; VI and VII need target redshift properties extracted from catalogs

### A.3.6. Use case 3.6

Show me a list of all data that satisfies:

    I. DataType=Cube with 3 dimensions
    II. Axes includes FREQ
    III. Axes includes RA
    IV. Axes includes DEC
    V. Frequency resolution < 10 MHz
    VI. Rest Frequency = 337.2966 GHz appears in band
    VII. Any observation that could have detected a line at this rest frequency from any target, using the nominal redshift for the target.

### A.3.7. Use case 3.7

Show me a list of all data that satisfies:

      I. DataType=Cube with 3 dimensions
      II. Axes includes FREQ
      III. Axes includes RA
      IV. Axes includes DEC
      V. Frequency resolution < 10 MHz
      VI. Rest Frequency in (213.36053, 256.0278, 298.6908925, 341.350826, 384.0066819, 426.6579505, 469.3041221, 511.944687, 554.5791355) GHz appears in band
      VII. Any observation that could have detected HCS+ (list of transition rest frequencies given above) from any target, using the nominal redshift for the target.

### A.3.8. Use case 3.8

Show me a list of all data that satisfies:

    I.   DataType=Cube with 4 dimensions
    II.  Axes includes FREQ
    III. Axes includes RA with > 100 pixels
    IV. Axes includes DEC with > 100 pixels
    V.  Axes includes STOKES
    VI. Frequency resolution < 1 MHz
    VII. Rest Frequency = 345.795990 GHz appears in band
         NB: Need for a polarisation axis

### A.3.9.   Use case 3.9

Looking for moving targets:

    I.   Show me the names of all the objects that have moving coordinates (i.e. no fixed RA, DEC position).

## A.4.  Discovering time series

### A.4.1.   Use case 4.1

Show me a list of all data which satisfies:

    I.   DataType=TimeSeries
    II.  RA includes 16.00 hours
    III. DEC includes +41.00
    IV. Time resolution better than 1 minute
    V.  Time interval (start of series to end of series) > 1 week
    VI. Observation data before June 10, 2008
    VII. Observation data after June 10, 2007

## A.5.  Discovering general data

### A.5.1.   Use case 5.1

Show me a list of all data that satisfies:

    I.   Optical imaging
    II.  In the M81 group
    III. With area greater than 0.5 square degrees
    IV. With sensitivity > 10 _ _ for point source m=25
    V.  I also want X-ray data with cutouts 5 arcmin on a side of all the detected galaxies
    VI. I also want Radio data cutouts 5 arcmin on a side around detected galaxies

### A.5.2.   Use case 5.2

Show me a list of all data that satisfies:

    I.   DataType=Imaging or Spectroscopy
    II.  RA includes 16.00 hours
    III. DEC includes +41.00 degrees
    IV. SDSS images and spectra AND CFHTLS images and spectra

### A.5.3.   Use case 5.3

In Virgo cluster show me imaging and X-ray data for all galaxies that are cluster members and have B < 21.

## *A.6. Other Use Cases*

### A.6.1. Use case 6.1

Given COSMOS (or other survey) X-Ray source catalogue give me all the sources with photoZ > X, and spiral galaxy counterpart and produce radio - to -X-ray SEDs.

Comment: Requires source/object catalogues to drive data query (for SED info which may be catalogue or data).

### A.6.2. Use Case 6.2

Given a list of Abell clusters, give me all their Chandra images with exposure time > X, after I select regions occupied by the diffuse emission, give me all the Chandra point sources in these regions, and find their redshift (I want to find background quasars because I am interested in lensing and I have no idea where to go to find z). For the quasars, give me high resolution (< 0.5") optical and radio images, and build SEDs.

*Comment*: Requires source/object catalogues and interactive image interactions (applications/interfaces), further query, and more catalogues to drive data query.

### A.6.3. Use case 6.3

Find me all the variable Chandra sources with optical counterpart and redshift. If redshift is not available, give me an SED to compare with source templates (I also would like to run a tool or obtain a library of such templates from a theory database, which I expect the VO to provide). My aim is to separate stars from variable quasars.

*Comment*: Complex use-case, including templates and theory as well as catalogues.

# Appendix B: ObsCore Data Model Detailed Description

This section provides a full description of all data model elements including both mandatory and optional elements (specified by the value in the "MAN" column). The full Utype for all elements of the Observation Core Components data model includes an "*obscore:*"prefix (defining the namespace for ObsCoreDM) which has been elided here for brevity.

| Column Name | Utype | Unit | Type | Description | MAN |
|---|---|---|---|---|---|
| OBSERVATION INFORMATION (section B.1) | | | | | |
| dataproduct_type | Obs.dataProductType | unitless | enum string | Data product (file content) primary type | YES |
| dataproduct_subtype | Obs.dataProductSubtype | unitless | string | Data product specific type | NO |
| calib_level | Obs.calibLevel | unitless | enum int | Calibration level of the observation: in {0, 1, 2, 3} | YES |
| TARGET INFORMATION (section B.2) | | | | | |
| target_name | Target.Name | unitless | string | Object of interest | YES |
| target_class | Target.Class | unitless | string | Class of the Target object as in SSA | NO |
| DATA DESCRIPTION (section B.3) | | | | | |
| obs_id | DataID.observationID | unitless | string | Internal ID given by the ObsTAP service | YES |
| obs_title | DataID.Title | unitless | string | Brief description of dataset in free format | NO |
| obs_collection | DataID.Collection | unitless | string | Name of the data collection | YES |
| obs_creation_date | DataID.Date | unitless | date | Date when the dataset was created | NO |
| obs_creator_name | DataID.Creator | unitless | string | Name of the creator of the data | NO |
| obs_creator_did | DataID.CreatorDID | unitless | string | IVOA dataset identifier given by the creator | NO |
| CURATION INFORMATION (section B.4) | | | | | |
| obs_release_date | Curation.releaseDate | unitless | string | Observation release date (ISO 8601) | NO |
| obs_publisher_did | Curation.PublisherDID | unitless | string | Dataset ID given by the publisher. | YES |
| publisher_id | Curation.PublisherID | unitless | string | IVOA-ID for the Publisher | NO |
| bib_reference | Curation.Reference | unitless | string | Service bibliographic reference | NO |
| data_rights | Curation.Rights | unitless | enum | Public/Secure/Proprietary/ | NO |
| ACCESS INFORMATION (section B.5) | | | | | |
| access_url | Access. Reference | unitless | string | URL used to access dataset | YES |
| access_format | Access. Format | unitless | string | Content format of the dataset | YES |
| access_estsize | Access.Size | kbyte | int | Estimated size of dataset: in kilobytes | YES |
| SPATIAL CHARACTERISATION (section B6.1) | | | | | |

| s_ra | Char.SpatialAxis.Coverage.Location.Coord.Position2D.Value2.C1 | deg | double | Central Spatial Position in ICRS Right ascension | YES |
|---|---|---|---|---|---|
| s_dec | Char.SpatialAxis.Coverage.Location.Coord.Position2D.Value2.C2 | deg | double | Central Spatial Position in ICRS Declination | YES |
| s_fov | Char.SpatialAxis.Coverage.Bounds.Extent.diameter | deg | double | Estimated size of the covered region as the diameter of a containing circle | YES |
| s_region | Char.SpatialAxis.Coverage.Support.Area | | AstroCoordArea | Region covered in STC or ADQL | YES |
| s_resolution | Char.SpatialAxis.Resolution.refval | arcsec | double | Spatial resolution of data as FWHM | YES |
| s_ucd | Char.SpatialAxis.ucd | unitless | string | Ucd for the nature of the spatial axis (pos or u,v data) | NO |
| s_unit | Char.SpatialAxis.unit | unitless | string | Unit used for spatial axis | NO |
| s_resolution_min | Char.SpatialAxis.Resolution.Bounds.Limits.Interval.LoLim | arcsec | double | Resolution min value on spatial axis (FHWM of PSF) | NO |
| s_resolution_max | Char.SpatialAxis.Resolution.Bounds.Limits.Interval.HiLim | arcsec | double | Resolution max value on spatial axis | NO |
| s_calib_status | Char.SpatialAxis.calibStatus | unitless | enum | Type of calibration along the spatial axis | NO |
| s_stat_error | Char.SpatialAxis.Accuracy.statError.refval.value | arcsec | double | Astrometric precision along the spatial axis | NO |
| **TIME CHARACTERISATION (section B6.3)** | | | | | |
| t_min | Char.TimeAxis.Coverage.Bounds.Limits.Interval.StartTime | d | double | Start time in MJD | YES |
| t_max | Char.TimeAxis.Coverage.Bounds.Limits.Interval.StopTime | d | double | Stop time in MJD | YES |
| t_exptime | Char.TimeAxis.Coverage.Support.Extent | s | double | Total exposure time | YES |
| t_resolution | Char.TimeAxis.Resolution.refval | s | double | Temporal resolution FWHM | YES |
| t_calib_status | Char.TimeAxis.calibStatus | unitless | enum | Type of time coord calibration | NO |
| t_stat_error | Char.TimeAxis.Accuracy.StatError.refval.value | s | double | Time coord statistical error | NO |
| **SPECTRAL CHARACTERISATION (section B6.2)** | | | | | |
| em_ucd | Char.SpectralAxis.ucd | unitless | string | Nature of the spectral axis | NO |
| em_unit | Char.SpectralAxis.unit | unitless | string | Units along the spectral axis | NO |
| em_calib_status | Char.SpectralAxis.calibStatus | unitless | enum | Type of spectral coord calibration | NO |
| em_min | Char.SpectralAxis.Coverage.Bounds.Limits.Interval.LoLim | m | double | start in spectral coordinates | YES |
| em_max | Char.SpectralAxis.Covera | m | double | stop in spectral coordinates | YES |

| | ge.Bounds.Limits.Interval.HiLim | | | | |
|---|---|---|---|---|---|
| em_res_power | Char.SpectralAxis.Resolution.ResolPower.refval | unitless | double | Value of the resolving power along the spectral axis. (R) | YES |
| em_res_power_min | Char.SpectralAxis.Resolution.ResolPower.LoLim | unitless | double | Resolving power min value on spectral axis | NO |
| em_res_power_max | Char.SpectralAxis.Resolution.ResolPower.HiLim | unitless | double | Resolving power max value on spectral axis | NO |
| em_resolution | Char.SpectralAxis.Resolution.refval.value | m | double | Value of Resolution along the spectral axis | NO |
| em_stat_error | Char.SpectralAxis.Accuracy.StatError.refval.value | m | double | Spectral coord statistical error | NO |
| OBSERVABLE AXIS (sectionB6.4) | | | | | |
| o_ucd | Char.ObservableAxis.ucd | unitless | string | Nature of the observable axis | YES |
| o_unit | Char.ObservableAxis.unit | unitless | enum | Units used for the observable values | NO |
| o_calib_status | Char.ObservableAxis.calibStatus | unitless | enum | Level of calibration for the observable coordinate | NO |
| o_stat_error | Char.ObservableAxis.Accuracy.StatError.refval.value | units specified by o_unit | double | Statistical error on the Observable axis | NO |
| POLARISATION AXIS (section B6.6) | | | | | |
| pol_states | Char.PolarizationAxis.stateList | unitless | string | List of polarization states measured in this data set | YES |
| PROVENANCE (section B7) | | | | | |
| facility_name | Provenance.ObsConfig.facility.name | unitless | string | [from the VODataService Standard] | YES |
| instrument_name | Provenance.ObsConfig.instrument.name | unitless | string | The name of the instrument used for the observation | YES |
| proposal_id | Provenance.Proposal.identifier | string | string | Identifier of proposal to which observation belongs | NO |

**Table 5**: Data model summary

## B.1. Observation Information

This class is a place holder that gathers all metadata relative to an observed and distributed dataset. It points to existing classes of Spectrum DM and types from VODataService (Plante & al., 2010) .

### B.1.1. Data Product Type *(dataproduct_type)*

The model defines a data product type attribute for the Observation Class. It is the type of the observation product the user queries for or selects for retrieval. Only high level generic types are defined at this level to allow global data discovery queries to be posed to multiple data archives. This is coded as a string that conveys a general idea of the content and organization of a dataset. Possible values for this string are described in section 3.3.1. The short name for this attribute is *dataproduct_type* in the implemented Ivoa.ObsCore table, as shown in Table 4.

### B.1.2. Data Product Subtype *(dataproduct_subtype)*

In order to be more precise the data product type may be refined with a second field, the data product *subtype*. Unlike the more generic *dataproduct_type*, this field is intended to precisely specify the scientific nature of the data product, possibly in terms relevant only to a specific archive or data collection. While less useful for global data discovery this allows the data products within a specific archive to be precisely identified and referenced in queries of that specific archive. The data provider should define a vocabulary sufficient to classify all science data products in their archive to be exposed with ObsTAP. In the future we may be able to define broader standards to classify data at this level, although it will likely always be the case that data differs at the level of specific instrumental survey data collections. The data product subtype allows data within a specific archive or data collection to be precisely classified and referenced in subsequent discovery queries. Subtype names should be simple identifiers or dot-delimited tokens (a, a.b, a.b.c, etc.).

### B.1.3. Calibration level *(calib_level)*

It is a convention we suggest to use to classify the different possible calibration status of an observed dataset. These 4 categories allow distinguishing 4 levels of calibration and would be sufficient for 80% of the data collections. This will be up to the data provider to consider how to map his own internal classification to the suggested scale here.

Following examples can help to find the most appropriate value for the *calibLevel* attribute (Obs.calibLevel).

> **Level 0**: Raw instrumental data, possibly in proprietary internal provider format, that needs specific tools to be handled.

> **Level 1**: Instrumental data in a standard format (FITS, VOTable, SDFITS, ASDM, etc. The data may or may not be calibrated. Standards tools can handle it.

> **Level 2**: Science ready data, with instrument signature removed, and calibration status defined on all physical axes.

> **Level 3**: Enhanced data products like mosaics, improved co-added image cubes, resampled or drizzled images, etc., spectra with calibrated velocity axis at a particular line rest frequency. In such case, the improved calibration procedure is described by the data provider in some way; progenitors of such a data product can be identified into the reduction pipeline.

This classification is simple enough to cover all regimes. Data providers will adjust the mapping of their various internal levels of calibration to this general frame.

## *B.2. Target*

This is the astronomical object of interest for which the observation was performed. Only the name is used in the Core Components model but the Target object is fully designed in the Spectrum data model. Serendipitous archives or surveys may not contain this information for all observations, so this can be 'NULL' if necessary or have a value like 'DARK' for instance to specify a dark field and not a pointed observation.

### B.2.1. Target Name *(target_name)*

The *target_name* specified here is often un-reliable when one is searching for specific classes of objects, at least for most archives since it is quite hard to standardize automatically the target name for each observation. So the users could be warned that specifying *target_name* for their search will not necessarily return the expected results. It is quite useful for moving target like planets.

### B.2.2. Class of the Target source/object *(target_class)*

This field indicates the type of object that was pointed for this observation. It is a string with possible values defined in a special vocabulary set to be defined: list of object classes (or types) used by the SIMBAD database, NED or defined in another IVOA vocabulary.

## *B.3. Dataset Description*

After acquisition and reduction an observation is uniquely identified by its creator and gets a creator dataset identifier. This information is defined in the Spectrum data model in the *DataID* class. We re-use this class and the Utype DataID.CreatorDID in order to distinguish two datasets curated by two different services (archives) but originating from the same creator. When broadcasting a query to multiple servers, the response may contain multiple copies of the same dataset, with a unique DataID.CreatorDID *obs_creator_did* but possibly different *obs_publisher_did* (given by the data provider). Therefore a unique identifier is needed here. In the ObsCore model, the short name associated to this ID should be *obs_id* (to be checked).

The second identifier used in this model is the one given by the data provider, and defined in the Curation Class.

### B.3.1. Creator name (*obs_creator_name*)

The name of the institution or entity which created the dataset, in a simple string.

### B.3.2. Observation Identifier *(obs_id)*

The *obs_id* column contains a collection-specific identifier for an observation. In the case where multiple data products are available for an observation (e.g. with different calibration levels), the *obs_id* value will be the same for each product of the observation. This is equivalent to the dataset name for many archives where dataset name could have many files associated with them.

### B.3.3. Dataset Text Description *(obs_title)*

This data model field re-uses a field from the Spectrum Data model: DataID.Title. It should contain a brief description (displayable in less than one line of text) specifying in scientific terms the content of the dataset. The contents of this field are free format and are up to the data provider. For example a radio survey field consisting of HI and CO cubes and an associated 2D continuum image might use *obs_title* to describe the individual data products as "HI cube", "CO cube", "Stokes I continuum image at 1420 MHz", and so forth.

This is commonly used in analysis software to e.g. describe a dataset in a query response table, in a plot header, in the label of a displayed image, and so forth. This helps the user to check the validity and pertinence of a selected data set for his/her personal goal.

### B.3.4. Collection name (*obs_collection*)

The name of the collection (DataID.Collection) identifies the data collection to which the data product belongs. A data collection can be any collection of datasets which are alike in some fashion. Typical data collections might be all the data from a particular telescope, instrument, or survey. The value is either the registered shortname for the data collection, the full registered IVOA identifier for the collection, or a data provider defined *shortname* for the collection. Examples: HST/WFPC2, VLT/FORS2, CHANDRA/ACIS-S, etc.

We understand that this is not a very precisely defined field. What is important for the data provider is to use the collection name which is meaningful to astronomers.

### B.3.5. Creation date (*obs_creation_date*)

The date when the dataset was created. This is a time stamp, stored in ISO 8601 format, using this specific format :("YYYY-MM-DDThh:mm:ss").

### B.3.6. Creator name (*obs_creator_name*)

The name of the institution or entity which created the dataset.

### B.3.7. Dataset Creator Identifier *(obs_creator_did)*

IVOA dataset identifier given by its creator. See definition in the SpectrumDM specification (McDowell, Tody, & al, 2011)

## B.4. Curation metadata

The *Curation* class inherits from the Spectrum data model and VOResource concepts too. The various attributes for ObsCore are:

### B.4.1. Publisher Dataset ID *(obs_publisher_did)*

This is the identifier the publisher provides for this observation. It may differ from the original identifier given by the creator of the dataset. (new reduction, new version, etc..). The corresponding Utype mapped from the Spectrum DM is *Curation.PublisherDID* and relates to the same definition.

This field contains the IVOA dataset identifier (Plante & al., 2007) for the published data product. This value must be unique within the namespace controlled by the dataset publisher (data center). It will also be globally unique since each publisher has a unique registered publisher ID. The same dataset may however have more than one publisher dataset identifier if it is published in more than one location (the creator DID, if defined for the given dataset, would be the same regardless of where the data is published).

### B.4.2. Publisher Identifier *(publisher_id)*

The IVOA ID for the data provider as defined in the Spectrum DM.

### B.4.3. Bibliographic Reference *(bib_reference)*

URL or bibcode for documentation. This is a forward link to major publications which reference the dataset. This is re-used from the SSA definition. See (Tody, Dolensky, & al., 2011) in section 4.2.5.6 about Curation Metadata.

### B.4.4. Data Rights (*data_rights*)

This parameter allows mentioning the availability of a dataset. Possible values are: *public*, *secure*, or *proprietary* as stated in the VODataService recommendation (Plante & al., 2010).

### B.4.5. Release Date (*obs_release_date*)

This is a new attribute added to the original Curation class inherited from the Spectrum Data Model.

It specifies the date of public release for an observation or a data product. This time stamp is a convenient way to distinguish public and private observations and also tell users when a specific data product will become available. The value is in ISO 8601 format reusing this pattern: ("YYYY-MM-DDThh:mm:ss") and could be NULL. An observation with a NULL value in the *releaseDate* attribute is proprietary by definition.

## B.5. Data Access

The data format as well as the URL to access the dataset is provided by the **Access** Class inherited from the SSA Utype list (Tody, Dolensky, & al., 2011). Also included is an attribute for the estimated size of the data file.

### B.5.1. Access Reference *(access_url)*

This item (Access.Reference) contains a URL that can be used to download the data product (as a file of some sort).

Users should be aware that these URL values could be volatile and subject to be different from one access time to another. The access reference URL may refer to a static object or may cause data to be generated on the fly, so long as access is synchronous.

### B.5.2. Access Format (access_format)

This data model item is defined in section 4.7 where you can find the list of possible values.

We are aware that many different domains and applications need to define data formats, and then define a controlled vocabulary based on implementation feedback given by data providers at different sites.

### B.5.3. Estimated Size (access_estsize)

The Access.Size field contains the approximate size (in kilobytes) of the file available via the *corresponding url*. This is used only to gain some idea of the size of a data product before downloading it, hence only an approximate value is required. It is only a useful indication that can help to tune download functionalities in an application according to high volumes of data and transfer bit rate.

## B.6. Description of physical axes: Characterisation classes

As mentioned in the use-cases, selection criteria for an observation depend on the physical axes contained in the dataset especially the position, band, time, and the type of observed quantity that we call "observable" in the data model. The **observable axis** can cover various types of flux but also velocity, etc. Such a description was tackled in the IVOA Characterisation data model (Louys & DataModel-WG., 2008) from which we re-use mainly the first level and second levels of details except for the spatial coverage where the support region (level 3) is used too.

### B.6.1. Spatial axis

### B.6.1.1. The observation reference position: *(s_ra and s_dec)*

Two coordinates in position are used to identify a reference position (typically the center) of an observation in the sky, attached to a coordinate system definition.

- Coordinate system

  The coordinate system defined in the Characterisation DM is based on the STC:Coordsys class. The model in principle supports all kind of coordinate systems defined in the STC reference list (Rots, 2007). However, the ObsTAP implementation of the model mandates that queries expressed in ICRS should be supported in an ObsTAP service. This allows a general query to be sent to multiple archives or data centers, but requires some interpretation /conversion of coordinates at the server side. Still this is efficient for the large data discovery strategy we need to provide. Data in a specific coordinate system will be available via client applications that would do the conversions or adapt the coordinate system to some specific servers.

- Coordinates

  The model uses the *Location* Class from the Characterisation DM, with the Utype values:

  *Char.SpatialAxis.Coverage.Location.Coord.Position2D.Value2.C1*

  *Char.SpatialAxis.Coverage.Location.Coord.Position2D.Value2.C2*

  whose short names in the ObsCore table are **s_ra** and **s_dec.** We assume that ObsTAP implements these coordinates in the ICRS system.

Using other coordinate systems as defined in STC (Rots, 2007) and re-used in the Characterisation DM can be considered in client applications in charge of the coordinate translations.

## B.6.1.2. The covered region

The Coverage class along the spatial axis provides two possible concepts:

- **Bounds** which in turn can use two representations:

  a) **A bounding box** that can estimate very coarsely the coverage of an observation.  It is modeled as a couple of intervals on each coordinates with Utypes:

  *Char.SpatialAxis.Coverage.Bounds.limits.Interval.LoLimit2Vec.C1*

  *Char.SpatialAxis.Coverage.Bounds.limits.Interval.HiLimit2Vec.C1*

  *Char.SpatialAxis.Coverage.Bounds.limits.Interval.LoLimit2Vec.C2*

  *Char.SpatialAxis.Coverage.Bounds.limits.Interval.HiLimit2Vec.C2*

  b) The **extent of the field** of view *(s_fov)*

  The model offers to estimate the size of the diameter of the greater circle encompassing the field of view.

This is not covered by the Characterisation DM v1.1 but in the new release of Characterisation v2.0 as *Char.SpatialAxis.Coverage.Bounds.Extent.diameter*, a new definition added in Characterisation DM v2.0 ( to appear).

- **Support**: (s_region)

  A precise region description of spatial footprint of the dataset using region types like Circle, Polygon, etc., provided in STC.  The Utypes:

  *Char.SpatialAxis.Coverage.Support.Area*
  *Char.SpatialAxis.Coverage.Support.AreaType*

  define this region, and STC-S can be used to serialise the values.

## B.6.1.3. Spatial Resolution (*s_resolution*)

The minimal size that can be distinguished along the spatial axis, *s_resolution* is specified in arcseconds and has the following Utype: *Char.SpatialAxis.Resolution.RefVal.*

When this value is difficult to evaluate or inadequate, the range of possible resolution can be given in the optional data model fields: *s_resolution_min* and *s_resolution_max, as shown in Table 7.*

## B.6.1.4. Astrometric Calibration Status: (*s_calib_status*)

A string to encode the calibration status along the spatial axis (astrometry).

Possible values could be **{uncalibrated, raw, calibrated}** and correspond to the Utype *Char.SpatialAxis.calibStatus*

For some observations, only the pointing position is provided (*s_calib_status* ="uncalibrated"). Some other may have a raw linear relationship between the pixel coordinates and the world coordinates (*s_calib_status* ="raw").

### B.6.1.5. Astrometric precision (*s_stat_error*)

This parameter gives an estimate of the astrometric statistical error after the astrometric calibration phase. The corresponding Utype is: *Char.SpatialAxis.Accuracy.StatError.Refval.*

### B.6.1.6. Spatial sampling (*s_pixel_scale*)

This corresponds to the sampling precision of the data along the spatial axis. It is stored as a real number corresponding to the spatial sampling period, i.e., the distance in world coordinates system units between two pixel centers. It may contain two values if the pixels are rectangular.

### B.6.2. Spectral axis

This axis is generally used to represent different kinds of physical measurements: wavelength, energy, frequency or some interpretation of this with respect to a reference position like velocity.

The data model distinguishes the various flavors of this axis using the UCD attached to it, *Char.SpectralAxis.ucd* named as ***em_ucd*** *in* ObsTAP optional fields. Possible values for this UCD are defined in the Spectrum DM (McDowell, Tody, & al, 2011) in section 4.1.

Depending on the UCD used to specify the axis, the ObsCore model allows to describe the spectral coordinates in a relevant unit, corresponding to the spectral quantity, and specified in the model in *Char.SpectralAxis.unit* (***em_unit***)

Here is a short list of preferred value for the Observation data model Core Components extracted from the recommended values proposed in the Spectrum DM.

| Spectral coordinate | Char.SpectralAxis.ucd | Char.SpectralAxis.unit |
|---------------------|------------------------|------------------------|
| frequency | em.freq | Hz |
| wavelength | em.wl | m or angstrom |
| energy | em.energy | keV, J, erg |

Note that for the ObsTAP implementation, the Spectral axis coordinates are constrained as a wavelength quantity expressed in meters as mentioned in section 4.17

### B.6.2.1. Spectral calibration status (*em_calib_status*)

This attribute of the spectral axis indicates the status of the data in terms of spectral calibration. Possible values are defined in the Characterisation Data Model and belong to **{calibrated, uncalibrated, relative, absolute}.**

### B.6.2.2. Spectral Bounds

These are the limits of the spectral interval covered by the observation, in short *em_min* and *em_max.*

These limiting values are compatible with definitions of the physical quantity defined in the ucd and unit fields.

*In the ObsTAP implementation* such values are expressed as wavelength but using meters as units, as it is easily convertible.

### B.6.2.3. Spectral Resolution

As in the Characterisation data model we distinguish a reference value of the point spread function along the spectral axis from the resolution power along this axis, more appropriate when the resolution varies along the spectral axis. Only one of the following is needed in the data model:

    a)  A reference value for **Spectral Resolution (em_resol)**

A mean estimate of the resolution, e.g. Full Half Width Maximum (FWHM) of the Line Spread Function (or LSF). This can be used for narrow range spectra whereas in the majority of cases, the resolution power is preferable due to the LSF variation along the spectral axis. The corresponding Utype is *Char.SpectralAxis.Resolution.refval.value.*

    b)  A reference value for **Resolving Power (em_res_power)**

This is an average estimation for the spectral resolution power stored as a double value, with no unit. *Char.SpectralAxis.Resolution.resolPower refval*

    c)  **Resolving Power** limits *(em_res_power_min, em_res_power_max)*

These parameters simply give the limits of variation of the resolution power in the observation as minimal and maximal values and use the following Utypes:

        Char.SpectralAxis.Resolution.resolPower.LoLim

        Char.SpectralAxis.Resolution.resolPower.HiLim

### B.6.2.4. Accuracy along the spectral axis *(em_stat_error)*

This is also provided in the Characterisation data model, using the item mapped to the Utype: *Char.SpectralAxis.Accuracy.StatError.refval.value* and, stored in the same units as all the other spectral quantities.

### B.6.3. Time axis

### B.6.3.1. Time coverage *(t_min, t_max, t_exptime)*

Three time stamps are used: *t_start*, *t_stop*, and *t_exptime* the exposure time.A format like MJD is useful for easy calculations and preferred for the Observation Core components model. Other information is given in subsection 4.14 and 4.15.

### B.6.3.2. Time resolution *(t_resolution)*

Estimated or average value of the temporal resolution with Utype

        *Char.TimeAxis.Resolution.refval.value*

### B.6.3.3. Time Calibration Status: *(t_calib_status)*

This parameter gives the status of time axis calibration. This is especially useful for time series.

Possible values are principally **{calibrated, uncalibrated, relative, raw}.** This may be extended for specific time domain collections.

### B.6.3.4. Time Calibration Error: (*t_stat_error*)

A parameter used if we can estimate a statistical error on the time measurements (for time series again).

### B.6.4. Redshift Axis:

In order to support queries looking for radial velocity measurements, we envision including the Redshift Axis description as presented in the Spectrum Data model.

The full description of the appropriate data model fields will be prepared and added in the next version of the ObsCore specification.

### B.6.5. Observable Axis:

### B.6.5.1. Nature of the observed quantity (*o_ucd*)

Most observations measure some flux quantity depending on position, spectral coordinate, or time. Here we consider a more general axis: the "observable axis" that can be either flux or any other quantity, the nature of which is specified by the UCD attached to this axis.

The possible UCD values are part of the UCD1+ vocabulary (Preite Martinez, Derriere, Delmotte, Gray, & al., 2007) . One can find simple flux classes like: *phot.flux, phot.flux.density, phot.count, phot.mag,* or more complex combinations such as: *phot.flux.density;phys.polarization.stokes*

Various possibilities have been gathered at the following URL:

http://www.ivoa.net/internal/IVOA/ObsTap/ListForObservable25Oct2010.pdf

which provides a (non-exhaustive) list of possible triplets (observable name, UCD, units) for various observables data providers may want to describe in their archive. The units used to encode values of the Observable quantity are specified in *Char.ObservableAxis.unit* and can be exposed in ObsTAP with the optional field *o_unit*. See examples of unit strings in the table mentioned above.

### B.6.5.2. Calibration status on observable (Flux or other) (*o_calib_status*)

This describes the calibration applied on the Flux observed (or other observable quantity) .

It is a string to be selected in **{absolute, relative, normalized, any}** as defined in the SSA specification  (Tody, Dolensky, & al., 2011)  in section 4.1.2.10.

This list can be extended or updated for instance using an extension mechanism similar to the definition of new UCDs in the IVOA process, following the feedback from implementations of ObsTAP services.

### B.6.6. Polarisation measurements (*o_ucd, pol_states*)

This covers the case when the observed flux was recorded for various states of a polarizer. Then the dataset can be: a set of images, a set of spectra, a set of spectral cubes with various polarisation flux at each data point, etc.. What differs is the nature of the observable and the list of possible polarization states recorded.

In this case, *o_ucd* should at least contain the substring '**phys.polarisation'** like in "*phot.flux.density;**phys.polarisation**"* . The polarisation measure can be specified with UCD strings like in "*phot.flux.density;phys.polarisation.Stokes.I"*, etc. as shown in the list of observable UCD cited above.

In order to distinguish various polarization states recorded in a dataset, we define a polarization axis. *Char.PolarisationAxis.stateList* contains the list of the various polarization modes present in the dataset.

In the Obs/TAP implementation the column name is **pol_states**.It is a mandatory field with NULL value allowed if no polarizarion applies. Otherwise it contains a list of polarization labels inspired from the FITS specification. See Table 7 in FITS WCS Paper 1 (Greisen & Calabretta, 2002) . Labels are combined using symbols from the {I Q U V RR LL RL LR XX YY XY YX POLI POLA} set and separated by a **/** character. A leading **/** character must start the list. It should be ordered following the above list, compatible with the FITS list table for polarization definition.

Then a query can be easily written like:

```
SELECT * WHERE pol_states LIKE '%Y%'
```

which brings back all polarization moments of type :Y XY YX YY

On the contrary,

```
SELECT * WHERE pol_states LIKE '%/Y/%'
```

selects only datasets containing Y polarization state.

See A. Richards IVOA Note for the context of polarization data (Richards & Bonnarel, 2010). The full description of polarization metadata is covered in the upcoming Characterisation data model v2.0   (Bonnarel, Chilingarian, & Louys, (in prep.))

### B.6.7.  Additional Parameters on Observable axis

When implementing an ObsTap service , the archive manager may need to publish some parameters not present in the current version of ObsCore 1.0.

As an example, the type of noise present in an observation is not modeled. It depends on the instrument, on the data collection and can be defined in an optional column name in the IVOA.Obscore table like:

| Column Name | Datatype | Size | Units | ObsCoreDM Utype | UCD | Princ. | Index | Std |
|---|---|---|---|---|---|---|---|---|
| o_stat_error_type | adql:VARCHAR | 20 | NULL | NULL | stat.error;meta.code | *1* | *0* | *0* |

Possible values of o_stat_error_type could be: {poisson, gauss, speckle,,..} .and mentioned in the description of additional columns  (See section 4.20 for more details) o_ stat_error _mean, o_ stat_error _sigma can be defined as the parameters for the Gaussian case o_ stat_error_poisson as the Poisson gain, etc.

In case of these optional fields, defined by the data provider, the Utype column in the ObsCore table has a NULL value.

## B.7. Provenance

Provenance contains a class to represent the entire Observing configuration used to acquire an observation.  Instrumental parameters are gathered here.

### B.7.1.  Facility (*facility_name*)

The Facility class codes information about the observatory or facility used to collect the data. In this model we define one attribute of Utype *Provenance.obsConfig.facility.name* which re-uses the Facility concept defined in the VODataService specification (Plante & al., 2010).

 For combined observations stemming from multiple facilities the name may contain a list of comma separated strings, or the word "Many"; if the list is too long, as defined in the VODataservice specification.

The definition of a list of possible name values could be a task for the IVOA Semantic working group, starting from the ADS list published at http://vo.ads.harvard.edu/dv/facilities.txt and enriched when necessary.

## B.7.2. Instrument name (*instrument_name*)

The name of the instrument used for the acquisition of the observation. It is given in the model *as Provenance.ObsConfig.instrument.name* and encoded as a string. The possible name values could be checked in coordination with the Semantic WG too. Multiple values are also allowed for complex observations as defined for **facility name**.

## B.7.3. Proposal (*proposal_id*)

Each proposal has an identifier attribute that can be used to collect all observations and data products related to the same proposal. The corresponding Utype will simply be *Proposal.identifier*.

[NB: Here is presented only a minimal set of information on the instrumental configuration. See future documents on Provenance data model.]

# Appendix C: TAP_SCHEMA tables and usage

## C.1.   Implementation Examples

Examples of the ObsTAP use-cases and ObsTAP Schema can be found at the following URL:

*http://www.cadc.hia.nrc.gc.ca/cvo/*

This page will be kept current as the ObsTAP standard evolves.

## C.1.1.   Implementing a package of multiple data products

This example shows how to describe a complex observation, referenced by its *obs_id* field and containing different data products, all packed together in an archive file.

For example, for High Energy data sets we could have as the table response of an ObsTAP query:

| obs_id | data product Type | data product Subtype | Calibration Level | Access Format | Title |
|--------|-------------------|----------------------|-------------------|---------------|-------|
| 123 | event | chandra.hrc.pkg | 1 | application/x-tar-gzip | Chandra ACS-XYZ observation package (event,refimage) |
| 123 | image | chandra.hrc.refimage | 2 | image/fits | ACS-XYZ reference image |
| 123 | image | chandra.hrc.preview | 2 | image/jpeg | Chandra ACS-XYZ preview image |
| 345 | event | rosat.foo.pkg | 1 | application/x-tar-gzip | Rosat observation package |

The subtype could in principle be more generic but will likely be instrument-specific for a level 1 data product.

The Title should concisely describe the data product, e.g., origin, instrument, ID, what it is (observation package, calibration, standard view, etc.).  The title string is what one normally wants to output on a displayed image or plot to identify to a human the data being shown. Its length is limited to one line of text.

## C.2.   List of data model fields in TAP_SCHEMA

TAP Schema (`TAP_SCHEMA.columns`) metadata for all mandatory and optional data model fields are given in the following tables.  We suggest using only lower case for all column names in the tables used to implement ObsTAP, in order to simplify queries against multiple database systems.

Utypes strings are easier to read in 'Camel case' that is why we recommend using these strings as written in the tables below, for all interactions with users and developers. These strings are produced and read in VOTable for instance and may be consumed by some applications.

For Utypes originating from the Spectrum Data model, we keep the original writing.

For Utypes created from the UML ObsCore model, we apply these rules:

- Attributes of a class start with a lower case letter (e.g. calibStatus)
- For classes referencing one other class, we use the name of the reference or role, and not the one of the pointed class.

The meaning of the various columns corresponds to the definitions of the TAP IVOA standard (Dowler, Tody, & Rixon, 2010). See section 2.6.3 for the description of columns attributes.

As a reminder, the last three columns are implementation oriented:

'principal': means that this item is of main importance, and for instance is recommended in a select or should be shown in first priority in a query response.

'std': means this column is defined by some IVOA standard as opposed to a customized metadata defined by a specific service.

'indexed': tells if this column can be used as table index to optimize queries. Possible values for each of these three fields are integers, with this convention: (0=false, 1=true).

**Table 6** `TAP_SCHEMA.columns` values for the mandatory fields of an ObsTAP table. All Utypes have the data model namespace prefix **"obscore:"** omitted in the table.

| Column Name | Datatype | Size | Units | ObsCoreDM Utype | UCD | Principal | Index | Std |
|---|---|---|---|---|---|---|---|---|
| dataproduct_type | adql:VARCHAR | TBD | NULL | Obs.dataProductType | meta.id | 1 | TBD | 1 |
| calib_level | adql:INTEGER | NULL | NULL | Obs.calibLevel | meta.code;obs.calib | 1 | TBD | 1 |
| obs_collection | adql:VARCHAR | TBD | NULL | DataID.Collection | meta.id | 1 | TBD | 1 |
| obs_id | adql:VARCHAR | TBD | NULL | DataID.observationID | meta.id | 1 | TBD | 1 |
| obs_publisher_did | adql:VARCHAR | TBD | NULL | Curation.PublisherDID | meta.ref.url;meta.curation | 1 | TBD | 1 |
| access_url | adql:CLOB | NULL | NULL | Access.Reference | meta.ref.url | 1 | 0 | 1 |
| access_format | adql:VARCHAR | NULL | NULL | Access.Format | meta.code.mime | 1 | 0 | 1 |
| access_estsize | adql:BIGINT | NULL | kbyte | Access.Size | phys.size;meta.file | 1 | 0 | 1 |
| target_name | adql:VARCHAR | TBD | NULL | Target.Name | meta.id;src | 1 | 0 | 1 |
| s_ra | adql:DOUBLE | NULL | deg | Char.SpatialAxis.Coverage.Location.Coord.Position2D.Value2.C1 | pos.eq.ra | 1 | 0 | 1 |
| s_dec | adql:DOUBLE | NULL | deg | Char.SpatialAxis.Coverage.Location.Coord.Position2D.Value2.C2 | pos.eq.dec | 1 | 0 | 1 |
| s_fov | adql:DOUBLE | NULL | deg | Char.SpatialAxis.Coverage.Bounds.Extent.diameter | phys.angSize;instr.fov | 1 | 0 | 1 |
| s_region | adql:REGION | NULL | | Char.SpatialAxis.Coverage.Support.Area | phys.angArea;obs | 1 | 0 | 1 |
| s_resolution | adql:DOUBLE | NULL | arcsec | Char.SpatialAxis.Resolution.refval | pos.angResolution | 1 | TBD | 1 |

| t_min | adql:DOUBLE | NULL | d | Char.TimeAxis.Coverage.Bounds.Limits.Interval.StartTime | time.start;obs.exposure | 1 | 0 | 1 |
|---|---|---|---|---|---|---|---|---|
| t_max | adql:DOUBLE | NULL | d | Char.TimeAxis.Coverage.Bounds.Limits.Interval.StopTime | time.end;obs.exposure | 1 | 0 | 1 |
| t_exptime | adql:DOUBLE | NULL | s | Char.TimeAxis.Coverage.Support.Extent | time.duration;obs.exposure | 1 | TBD | 1 |
| t_resolution | adql:DOUBLE | NULL | s | Char.TimeAxis.Resolution.refval | time.resolution | 1 | 0 | 1 |
| em_min | adql:DOUBLE | NULL | m | Char.SpectralAxis.Coverage.Bounds.Limits.Interval.LoLim | em.wl;stat.min | 1 | 0 | 1 |
| em_max | adql:DOUBLE | NULL | m | Char.SpectralAxis.Coverage.Bounds. Limits.Interval.HiLim | em.wl;stat.max | 1 | 0 | 1 |
| em_res_power | adql:DOUBLE | NULL | NULL | Char.SpectralAxis.Resolution.ResolPower.refVal | spect.resolution | 1 | TBD | 1 |
| o_ucd | adql:VARCHAR | TBD | NULL | Char.ObservableAxis.ucd | meta.ucd | 1 | 0 | 1 |
| pol_states | adql:VARCHAR | TBD | NULL | Char.PolarizationAxis.stateList | meta.code;phys.polarization | 1 | 0 | 1 |
| facility_name | adql:VARCHAR | NULL | NULL | Provenance.ObsConfig.facility.name | meta.id;instr.tel | 1 | TBD | 1 |
| Instrument_name | adql:VARCHAR | NULL | NULL | Provenance.ObsConfig.instrument.name | meta.id;instr | 1 | TBD | 1 |

**Table 7** `TAP.schema.columns` values for the optional fields for an ObsTAP table. All Utypes have the data model namespace prefix **"obscore:"** omitted in the table.

| Column Name | Datatype | Size | Units | ObsCoreDM Utype | UCD | Principal | Index | Std |
|---|---|---|---|---|---|---|---|---|
| dataproduct_subtype | adql:VARCHAR | TBD | NULL | Obs.dataProductSubtype | meta.id | 1 | TBD | 1 |
| target_class | adql:VARCHAR | TBD | NULL | Target.Class | src.class | 1 | TBD | 1 |
| obs_creation_date | adql:TIMESTAMP | NULL | NULL | DataID.Date | time;meta.dataset | 1 | TBD | 1 |
| obs_creator_name | adql:VARCHAR | TBD | NULL | DataID.Creator | meta.id | 1 | TBD | 1 |
| obs_creator_did | adql:VARCHAR | TBD | NULL | DataID.CreatorDID | meta.id | 0 | TBD | 1 |
| obs_title | adql:VARCHAR | 200 | NULL | DataID.Title | meta.title;obs | 1 | 0 | 1 |
| publisher_id | adql:VARCHAR | TBD | NULL | Curation.PublisherID | meta.ref.url;meta.curation | 1 | TBD | 1 |
| bib_reference | adql:VARCHAR | TBD | NULL | Curation.Reference | meta.bib.bibcode | 0 | 0 | 1 |
| data_rights | adql:VARCHAR | NULL | NULL | Curation.Rights | meta.code | 0 | 0 | 1 |
| obs_release_date | adql:TIMESTAMP | NULL | NULL | Curation.releaseDate | time.release | 1 | 0 | 1 |
| s_ucd | adql:VARCHAR | NULL | NULL | Char.SpatialAxis.ucd | meta.ucd | 1 | 0 | 1 |
| s_unit | adql:VARCHAR | NULL | NULL | Char.SpatialAxis.unit | meta.unit | 1 | 0 | 1 |
| s_resolution_min | adql:DOUBLE | NULL | arcsec | Char.SpatialAxis.Resolution.Bounds.Limits.Interval.LoLim | pos.angResolution;stat.min | 1 | 0 | 1 |
| s_resolution_max | adql:DOUBLE | NULL | arcsec | Char.SpatialAxis.Resolution.Bounds.Limits.Interval.HiLim | pos.angResolution;stat.max | 1 | 0 | 1 |
| s_calib_status | adql:VARCHAR | NULL | NULL | Char.SpatialAxis.calibStatus | meta.code.qual | 1 | 0 | 1 |
| s_stat_error | adql:DOUBLE | NULL | arcsec | Char.SpatialAxis.Accuracy.statError.refval.value | stat.error;pos.eq | 0 | 0 | 1 |
| t_calib_status | adql:VARCHAR | NULL | NULL | Char.TimeAxis.calibStatus | meta.code.qual | 0 | 0 | 1 |
| t_stat_error | adql:DOUBLE | NULL | s | Char.TimeAxis.Accuracy.StatError.refval.value | stat.error;time | 0 | 0 | 1 |
| em_ucd | adql:VARCHAR | NULL | NULL | Char.SpectralAxis.ucd | meta.ucd | 1 | 0 | 1 |
| em_unit | adql:VARCHAR | NULL | NULL | Char.SpectralAxis.unit | meta.unit | 1 | 0 | 1 |
| em_calib_status | adql:VARCHAR | NULL | NULL | Char.SpectralAxis.calibStatus | meta.code.qual | 0 | 0 | 1 |
| em_res_power_min | adql:DOUBLE | NULL | NULL | Char.SpectralAxis.Resolution.ResolPower.LoLim | spect.resolution;stat.min | 1 | 0 | 1 |
| em_res_power_max | adql:DOUBLE | NULL | NULL | Char.SpectralAxis.Resolution.ResolPower.HiLim | spect.resolution;stat.max | 1 | 0 | 1 |
| em_resolution | adql:DOUBLE | NULL | m | Char.SpectralAxis.Resolution.refval.value | spect.resolution;stat.mean | 1 | 0 | 1 |
| em_stat_error | adql:DOUBLE | NULL | m | Char.SpectralAxis.Accuracy.StatError.refval.value | stat.error;em | 0 | 0 | 1 |
| o_unit | adql:VARCHAR | NULL | NULL | Char.ObservableAxis.unit | meta.unit | 1 | 0 | 1 |
| o_calib_status | adql:VARCHAR | NULL | NULL | Char.ObservableAxis.calibStatus | meta.code.qual | 1 | TBD | 1 |
| o_stat_error | adql:DOUBLE | NULL | o_unit | Char.ObservableAxis.Accuracy.StatError.refval.value | stat.error;phot.flux | 0 | 0 | 1 |
| proposal_id | adql:VARCHAR | NULL | NULL | Provenance.Proposal.identifier | meta.id; obs.proposal | 0 | TBD | 1 |